\documentclass[aps,preprint, showpacs]{revtex4}
\input{psfig.sty}

\usepackage{amsmath,bm}
\newcommand{\beq}{\begin{eqnarray}}
\newcommand{\eeq}{\end{eqnarray}}

\newcommand{\s}{\bm{\sigma}}

\newcommand{\J}{\bm{J}}

\newcommand{\C}{\bm{S}}
\newcommand{\K}{\mathbf{K}}

\newcommand{\p}{\bm{p}}

\newcommand{\hp}{\bm{\widehat{\p}}}

\newcommand{\x}{\bm{x}}

\newcommand{\0}{\bm{0}}

\newcommand{\bv}{\bm{\varphi}}

\newcommand{\hbv}{\bm{\widehat\varphi}}

\newcommand{\bl}{\bm{\lambda}}

\newcommand{\br}{\bm{\rho}}

\begin{document}

\title{Quantum states of indefinite spins: From baryons to massive
gravitino}

\author{M. Kirchbach}\email{mariana@ifisica.uaslp.mx}

\affiliation{Instituto de Fis\'{\i}ca, \\ 
Univ.\ Aut.\ de San Luis Potos\'{\i}, 
Av.\ Manuela Nava 6,
San Luis Potos\'{\i}, S.~L.~P.~78240,  M\'exico}

\date{\today}

\begin{abstract}
One of the long standing problems in particle physics is the covariant
description of higher spin states. The standard formalism is based upon 
totally symmetric Lorentz invariant tensors of rank-K with Dirac
spinor components, $\psi_{\mu_1...\mu_K}$, which satisfy the Dirac equation
for each space time index. In addition, one requires 
$p^{\mu_1}\psi_{\mu_1...\mu_K}=0$, and 
$\gamma^{\mu_1}\psi_{\mu_1...\mu_K}=0$. The solution obtained this way
(so called Rarita-Schwinger framework) describes the ``has--been'' 
spin-$(K+{1\over 2})$ particles in the rest frame, and particles
of uncertain ({\it fuzzy\/}) spin elsewhere. 
Problems occur when $\psi_{\mu_1...\mu_K}$
constrained this way
are placed within an electromagnetic field. In this case, the energy
of the has-been spin-$(K+{1\over 2})$ state becomes imaginary
(Velo-Zwanziger problem).
Here I consider two possible avenues for avoiding the above
problems. First I make the case that specifically for
baryon excitations there seems to be no urgency so far  for a 
formalism that describes isolated higher-spin states as all 
the observed nucleon and $\Delta (1232)$ excitations 
(up to $\Delta (1600)$) are exhausted by
{\em unconstrained\/} $\psi_\mu $, $\psi_{\mu_1...\mu_3}$, and 
$\psi_{\mu_1...\mu_5}$, which originate from rotational and
vibrational excitations of an underlying quark--diquark string.  
Second, I show that the
$\gamma^{\mu_1}\psi_{\mu_1...\mu_K} (\p ) =0$ constraint
is a short-hand of:
$-{1\over {2K+1}}\left({1\over m^2}W^2 +(K^2-{1\over 4})\right)
\psi_{\mu_1...\mu_K}= \psi_{\mu_1...\mu_K} $,
the covariant  definition of the unique
invariant subspace of the squared Pauli-Lubanski vector, $W^2$,
that is a parity singlet and of highest spin-$(K+{1\over 2})$ at rest.

I consider the simplest case of $K=1$ and
suggest to work in the sixteen dimensional vector
space, $\Psi$, of the direct product of the four-vector, $A_\mu$,
with the Dirac spinor, $\psi$,
i.e.  $\Psi =A \otimes \psi  $, rather than keeping 
space-time and spinor indices separated and show
that the ``has--been'' spin-3/2 piece is uniquely described
by means of the second order equation 
$\left(- {1\over 3}\left({1\over m^2}W^2 +{3\over 4}\right)-1 \right)\Psi =0$ 
without invoking any further supplementary 
conditions. In gauging the latter equation minimally and, in calculating
the determinant, one obtains the energy-momentum dispersion relation.
The latter turned out to be free from pathologies,
thus avoiding the classical Velo-Zwanziger problem.
\end{abstract}


\maketitle

\def\beq{\begin{eqnarray}}
\def\eeq{\end{eqnarray}}

\def\A{{\mathcal A}^\mu}
\def\W{{\mathcal W}_\mu}

\def\beq{\begin{eqnarray}}
\def\eeq{\end{eqnarray}}


\def\s{\mbox{\boldmath$\displaystyle\mathbf{\sigma}$}}
\def\Sg{\mbox{\boldmath$\displaystyle\mathbf{\Sigma}$}}
\def\J{\mbox{\boldmath$\displaystyle\mathbf{J}$}}
\def\K{\mbox{\boldmath$\displaystyle\mathbf{K}$}}
\def\P{\mbox{\boldmath$\displaystyle\mathbf{P}$}}
\def\p{\mbox{\boldmath$\displaystyle\mathbf{p}$}}
\def\hp{\mbox{\boldmath$\displaystyle\mathbf{\widehat{\p}}$}}
\def\x{\mbox{\boldmath$\displaystyle\mathbf{x}$}}
\def\0{\mbox{\boldmath$\displaystyle\mathbf{0}$}}
\def\bv{\mbox{\boldmath$\displaystyle\mathbf{\varphi}$}}
\def\hbv{\mbox{\boldmath$\displaystyle\mathbf{\widehat\varphi}$}}

\def\bl{\mbox{\boldmath$\displaystyle\mathbf{\lambda}$}}
\def\bl{\mbox{\boldmath$\displaystyle\mathbf{\lambda}$}}
\def\br{\mbox{\boldmath$\displaystyle\mathbf{\rho}$}}
\def\1{\openone}
\def\bfhh{\mbox{\boldmath$\displaystyle\mathbf{(1/2,0)\oplus(0,1/2)}\,\,$}}

\def\mn{\mbox{\boldmath$\displaystyle\mathbf{\nu}$}}
\def\amn{\mbox{\boldmath$\displaystyle\mathbf{\overline{\nu}}$}}

\def\mne{\mbox{\boldmath$\displaystyle\mathbf{\nu_e}$}}
\def\amne{\mbox{\boldmath$\displaystyle\mathbf{\overline{\nu}_e}$}}
\def\rlh{\mbox{\boldmath$\displaystyle\mathbf{\rightleftharpoons}$}}

\def\wm{\mbox{\boldmath$\displaystyle\mathbf{W^-}$}}
\def\hh{\mbox{\boldmath$\displaystyle\mathbf{(1/2,1/2)}$}}
\def\h00h{\mbox{\boldmath$\displaystyle\mathbf{(1/2,0)\oplus(0,1/2)}$}}
\def\znbb{\mbox{\boldmath$\displaystyle\mathbf{0\nu \beta\beta}$}}



\section{Particle states--revisited}
The definition of particle states is at the very heart 
of contemporary quantum field theories.
It is supposed to take its origin from frame-independent 
Casimir invariants of the Poincar\'e group as
was first noticed by Wigner in his work of late thirties.$^1$ 
Quantum states of free particles in a Poincar\'e
covariant framework have been considered
to transform for different inertial observers as
\begin{equation}
\Psi' (x)   = \exp \Big[ i 
( \epsilon^\mu P_\mu - \theta^{\mu \nu}M _{\mu \nu } )\,
\Big]\,   
\Psi (x)\, .
\label{state}
\end{equation}
Here,  $P_\mu $ and the totally antisymmetric 
tensor $M_{\mu\nu}$ with $\mu (\nu)=0,1,2,3$, are
the generators of the Poincar\'e group which satisfy the  
well known Poincar\'e algebra:
\begin{eqnarray}
\lbrack M_{\mu\nu},M_{\rho \sigma }\rbrack &=&
i\left( g_{\nu\rho }M_{\mu\sigma} - g_{\mu\nu }M_{\rho\sigma}
+g_{\mu\sigma}M_{\nu\rho} -g_{\nu\sigma}M_{\mu\rho}\right)\, ,\\
\lbrack P_\mu ,M_{\rho\sigma }\rbrack =i\left(
g_{\mu\rho} P_\sigma -g_{\mu\sigma}P_\rho\right)\, ,&& 
\lbrack P_\mu,P_\nu\rbrack =0\, ,\quad
M_{\mu\nu}=S_{\mu\nu}+ix_\mu\partial_\nu -ix_\nu\partial_\mu \, ,
\label{Po_alg}
\end{eqnarray}
where $\epsilon^\mu $ and $\theta^{\mu\nu}$ are continuous parameters,
$g_{\mu\nu}$=diag$(1,-1,-1,-1)$ is the metric tensor,
while  $S_{\mu\nu}$ is the purely intrinsic part
of $M_{\mu\nu}$ which later on will be associated with 
spin. In the standard convention, 
$P_\mu$ are the generators of the translation group,
${\mathcal T}_{1,3}$, in 1+3 time-space dimensions,
while  $M_{\mu \nu }$  are related
to the generators of boosts $(K_x$, $K_y$, $K_z$) and rotations
($J_x$, $J_y$, $J_z$) of the Lorentz group, $SO(1,3)$, via
\beq
M_{01}=K_x, \quad M_{02}= K_y, &\quad & M_{03}=K_z\, ,\nonumber\\
M_{12}=J_z, \quad M_{13}=-J_y\, , &\quad & M_{23}=J_x\, .
\label{Lor_gen}
\eeq
The Poincar\'e algebra has two invariant (Casimir) operators. 
These are the squared four-momentum, $P^2$, on the one side, and the 
squared Pauli-Lubanski vector, $W^2$, on the other side.
The Pauli--Lubanski vector is defined as$^2$ 
\begin{equation}
W_\mu  = -{1\over 2} 
\epsilon_{\mu\nu\rho\tau} S^{\nu\rho } P^\tau\, ,
\label{PauLu}
\end{equation}
where $\epsilon_{0123}=1$.
In terms of boost- and rotation generators,
the Pauli-Lubanski vector for states characterized
by the three momentum $\p$, is expressed as
\beq
W_\mu =\Big(-\C\cdot \p, -\C E +(\K\times \p)\Big)\, ,
\label{w_JK}
\eeq
Its squared (in covariant form) is calculated to be
\begin{equation}
W^2= 
- {1\over 2} S_{\mu\nu }S^{\mu\nu }\, P^2
+ S_{\mu\nu}P^\mu S^{\sigma\nu }  P_\sigma \, .
\label{w2}
\end{equation}
Note that for pure spin $(s,0)\oplus (0,s)$ spaces it can be shown
that $W^2$ simplifies to
\beq
W^2=-{1\over 4}S^{\mu\nu}S_{\mu\nu} \, P^2 \, .
\label{W2_j00j_3}
\eeq
In terms of Poincar\'e group invariants, the particle state definition
accepted so far in the literature prescribes that particles must have 
definite masses, $m$, and spins, $s$, according to
\beq
P^2\, \Psi (x)  &=& m^2  \Psi (x) \, ,\nonumber\\
W^2 \Psi (x) &=& - s(s+1) P^2 \Psi (x) \, .
\label{state_qn}
\eeq 
The latter equation imposes an essential
restriction onto the representation spaces of the
Poincar\'e group. In one of the possibilities, one can follow
Wigner and take the view that
particle states transform as {\it classical\/} unitary, and 
therefore infinite-dimensional, representations. 
However, according to Weinberg,$^3$ 
also {\it quantized\/} non-unitary finite-dimensional representations of 
the Lorentz group of the type $(s,0)\oplus (0,s)$ 
can be given particle interpretation because of unitarity
of the corresponding particle creation and annihilation operators. 
The persisting particle definitions deny, therefore,
existence of states without a definite mass or a definite spin.

The first quantum state to step out of the line 
was the physical neutrino of a particular flavor, $f$.
It is well known that such a neutrino ceases to have a well defined mass, 
a peculiarity that is manifest through the phenomenon of
neutrino oscillations,$^4$ a process
of crucial importance for the baryogeneses in the Universe,
\beq
|\nu_ f\rangle = \sum_{i=1}^{3}\, U_{fi}\, |m_i\rangle\, . 
\label{fuzzy_mass}
\eeq
Here, the mass eigenstates, $|m_i\rangle $, although possible,
serve solely as basis states as they are of indefinite flavor.
The unitary 3$\times $3 matrix $U_{fi}$ is the well known mixing
matrix.
At same time, in baryonic spectra, one sees resonances
of different spins and parities coming together to narrow mass bands
(see Figs. 1 and 2). In a recent analysis in Ref.~[6] 
I argued that such crops fit exactly into finite dimensional
non-unitary representations of the Lorentz group of
the type $\left({K\over 2},{K\over 2}\right)\otimes 
\Big[\left({1\over 2},0\right)\oplus \left(0,{1\over 2}\right)\Big]$ that are
described (in the momentum space of interest)
by means of a totally symmetric $K$-rank Lorentz tensor 
with Dirac spinor components, $\psi_{\mu_1...\mu_K}(\p )$.   
The quantum number $K$  is associated with
a Casimir invariant of the Lorentz--  (rather than the Poincar\'e) algebra,
which is determined as (see Ref.~[7]) 
${\mathcal C}\left( su(2)_L\oplus su(2)_R \right)={1\over 4} (\C^2-\K^2)$.
Its action upon $\psi_{\mu_1...\mu_K}(\p )$ amounts to,
\beq
{\mathcal C}\left( su_L(2)\oplus su(2)_R \right)\psi_{\mu_1...\mu_K}(\p )=
{K\over 2} ({K\over 2}+1)\psi_{\mu_1...\mu_K}(\p )\, .
\label{Lor_Cas}
\eeq
At rest, such states describe a
family of mass degenerate spin-parity states. For concreteness,
I  here consider the case of $K=3$ and unnatural parities
(to be of interest in the following) where
one finds the following spin- and parity sequence
\beq
\psi_{\mu_1...\mu_3}(\p )\stackrel{\mbox{rest}}{\longrightarrow}
{1\over 2}^-; {1\over 2}^+,{3\over 2}^+;{3\over 2}^-,{5\over 2}^-;
{5\over 2}^+, {7\over 2}^+\, .
\label{cluster_rest}
\eeq 
It is in particular interesting to study the action of the 
squared Pauli-Lubanski vector onto such spaces. In what follows
I will show that $W^2$ splits $\psi_{\mu_1...\mu_K}(\p )$ into
$(K+1)$ invariant subspaces. 

To begin with I first consider the case
of the rest frame where $W^2$ is particularly simple and
equals  $- \C^2 m^2$. In order to characterize the $W^2$ 
invariant subspaces it is favorable to introduce the additional
index $\tau_l^\pm  $ that takes the values 
$\tau_l^\pm = \left(l \pm {1\over 2}\right)\left( l \pm {1\over 2} +1 \right)$,
and $l=K, K-1, K-2, ..., 0$. 
Notice, that $\tau_l^+=\tau_{l+1}^-$.
In the following I introduce the compact notation
$\tau_l$ as $\tau_l:=\tau_l^+=\tau_{l+1}^-$.  As long states
differing by one unit in $l$ are of opposite parities, one 
finds the number of the parity degenerate 
$W^2$ invariant subspaces to equal $K$ according to
\beq
W^2\psi_{\mu_1...\mu_K}^{\tau_l}(\p ) =
- \tau_l\,  m^2 
\, \psi_{\mu_1...\mu_K}^{\tau_l}(\p )\, .
\label{W2_rest_sectors}
\eeq 
The latter are constituted of resonances having $l=0,1,..., K-1$.
There is only one parity singlet state, and it has $l=K$.
It is defined as $W^2\psi_{\mu_1...\mu_K}^{\tau^+_K}(\p ) =
- \tau^+_K\,  m^2 
\, \psi_{\mu_1...\mu_K}^{\tau^+_K}(\p )\, $.
The $W^2$ eigenvalues are frame independent,
and can be used to label the $W^2$ invariant subspaces in all 
inertial frames. 

Here, two very peculiar properties
of massive $\psi_{\mu_1...\mu_K}(\p )$ spaces show up.
To see them, one has to go back to $\left({K\over 2},{K\over 2}\right)$. 
The simplest one,
with $K=1$, was considered extensively in Ref.~[8].
Massive $\left({1\over 2},{1\over 2}\right)$'s 
(generically denoted by $A_\mu (\p )$) 
are associated with gauge bosons in
theories with spontaneously broken local gauge symmetries. 
Such representation spaces are spanned by four basis vectors. 
Three of them are divergence-less.
These are the two transversal degrees of freedom (d.o.f.) 
giving rise to left-- and right-handed circularly polarized gauge bosons,
on the one side, and 
the longitudinal polarization vector, on the other.  
Finally, the divergence-full time-like degree of freedom
is necessary to ensure completeness within the space under consideration.

\begin{enumerate}
\item  
\underline{The expel of the time-like degree of freedom from 
$\left({1\over 2},{1\over 2}\right)$ }:\quad
All massive gauge theories, be they Abelian or non-Abelian, are based
upon Proca's equation, $p^\mu A_\mu (\p )=0$. In so doing
one expels the time-like polarization vector 
which  seems {\it favorable\/}  because its norm, in being  
of sign opposite to one of the three remaining d.o.f., 
brings in unwanted imaginary masses after quantization.
Amazingly, at a later stage, the isolated degree of freedom has to 
reenter the theory through the back-door, in order
to  provide the St\"uckelberg term to the propagator of massive 
gauge bosons as required for renormalizability.$^9$
Nonetheless, physical reality at classical level is
still well designed by means of Proca's equation alone
in so far as  the time-like degree of freedom does not
contribute to the forces, i.e. to gradients of the gauge fields,
an observation first reported in Ref.~[10].
The lesson to be learned from the above considerations 
can be formulated as following.
Whenever we tailor a ``slim cut'' for a representation space 
with the conviction to better serve physical reality at 
the classical level, the expelled degree of freedom leave its 
footprint at the quantum level in form of troublesome divergences 
disturbing renormalization. It is at that level that the disregarded
degree of freedom has to be brought back to existence in terms
of somewhat artificially furnished renormalization schemes.

\item \underline{Indefinite spin in $\left({1\over 2},{1\over 2}\right)$}:
\quad
None of the four degrees of freedom within the massive
$\left({1\over 2},{1\over 2}\right)$ carries a definite intrinsic 
spin. This is so because the commutator between $W^2$ and 
$\C^2$ does not vanish co-variantly and, in effect,
$W^2$ invariant subspaces in (1/2,1/2) are no longer  
$\C^2$ eigenstates. 
In Ref.~[11], we calculated $[W^2,\C^2]=-4i E \K\cdot \p$.
In order to distinguish  $W^2$ for $(s,0)\oplus(0,s)$
representations, from $W^2$ for 
$({K\over 2},{K\over 2})$ representations, 
we introduced in Ref.~[11] the notion $\widetilde{W}^2$.
After all, in general, $\psi_{\mu_1...\mu_K}(\p )$ stand as examples
for states of indefinite ({\it fuzzy\/}) spins.
\end{enumerate}

Therefore, with the physical neutrino of a fixed flavor, a 
particle of {\it fuzzy\/} mass,
and the baryon resonances of the type $\psi_{\mu_1...\mu_K}(\p )$, 
particles of {\it fuzzy\/} spin, we here have at hand two 
contra-examples to what a particle should 
be according to trade-mark  definitions.
And yet, one can not deny particle status neither to  neutrino, 
nor to  baryon resonances. In this sense, extending
the particle definition seems inevitable. 
The extension can be such as to allow for the
possibility to use besides the Poincar\'e invariants
$p^2$ and $W^2$ also Lorentz covariant quantum numbers for 
labeling the states.
{}For example, $\psi_{\mu_1...\mu_K}(\p )$ could be primarily characterized by
$K$ before getting down to $\tau^\pm _l$ 
(see Section 4 below for more details).\\

Before proceeding further, a comment on the impact of the time-like
degree of freedom in $\left({1\over 2},{1\over 2}\right)$ 
(here denoted by $\epsilon_4 (\p )$)
onto the fermionic degrees of freedom
in the product space of the four-vector spinor $\psi_\mu (\p )$ is in order.
In momentum space, the fermionic degrees of freedom that take
their origin from $\epsilon_4 (\p )$ are 
$\epsilon _4 (\p) u_h(\p )$, and $\epsilon_4 (\p ) v_h(\p )$, respectively, 
where $u_h(\p )$ and $v_h(\p )$ are in turn Dirac's 
particle- and anti-particle spinors of momentum $\p$ and helicity $h$.
Because of the opposite sign of the norm of $\epsilon_4 (\p )$ relative
to the norm of the remaining three divergenceless basis vectors 
(here denoted by $\epsilon_i (\p )$ with $i=1,2,3$), 
one encounters the situation that if, say, $\epsilon_i (\p ) u_h(\p )$ are to
be associated with particles, $\epsilon _4 (\p ) u_h(\p )$ has to be a
associated with an anti-particle. Thus it looks like $\psi_\mu (\p )$
joins particles of opposite fermionic numbers which is likely to give rise to 
inconsistencies during propagation.
The way out of this is the following. 

First one has to clearly single out
the case of totally neutral massive $\psi_\mu (\p ) $ fields, such 
like the massive gravitino. In contrast to matter fermions,
gauge fermions can not be endowed by any fermion
numbers (charges) that need to be conserved.
Gravitino and anti-gravitino should be identical and not
distinct neither through opposite charges, nor through opposite
parities, contrary to usual fermions.
This is the price one has to pay for the
unrestricted back and forth exchange of fermions 
among other fields, where the vertex form should not change
in depending on
whether the gauge fermion under consideration
has been emitted as a particle, or, absorbed as an anti-particle.
Therefore, one is no longer interpretation bound to associate 
$u_h(\p )$ with particles, while $v_h (\p )$ with anti-particles.
Stated differently,  $\epsilon_4 (\p ) v_h(\p )$ is as good a particle
as is  $\epsilon _i (\p ) u_h(\p )$.
Therefore, massive gravitino presents itself as a $\psi_\mu (\p)$
wholeness of sixteen physical degrees of freedom, an observation 
already reported in our previous work in Ref.~[11].

Next, if one is to consider, say, the N(1440) state as a part
of the baryonic $\psi_\mu (\p )$, then one indeed would have to
describe it in terms of $\epsilon_4 (\p ) v_h(\p )$.
Simultaneously, $N(1520)$ and $N(1535)$ would require  
$\epsilon_i (\p ) u_h (\p )$.
In order to avoid confusion with the baryon number conservation, one
could look  on $\epsilon _i(\p )\psi (\p )$,
and $\epsilon_4 (\p ) \psi (\p )$ from the following 
independent perspectives. 
One can use
\beq
(p^\nu\gamma_\nu -m) u_\mu (\p )&=&0\, , 
\label{Dirac_RS_1}\\
p^\mu u _\mu (\p )&=&0\, ,
\label{Proca_SC}
\eeq
to pick up the ($D_{13}$-- $S_{11}$) sub-cluster from $\psi_\mu (\p )$.
In Refs.~[6] the ($D_{13}$-- $S_{11}$)
cluster propagator was given as
\beq
{\mathcal S}^{D_{13}S_{11}}_{\mu\nu }=
{{(-g_{\mu\nu} + {{p_\mu p_\nu}\over m^2})(p^\lambda\gamma_\lambda +m)}
\over {2m(p^2-m^2)}}\, .
\label{DS_prop}
\eeq
The $P_{11}$ (Roper) resonance needs to be treated independently, say,
by exploiting  once again
Dirac's equation, however, this time supplemented by a different
auxiliary condition that removes the parity degeneracy of
the $W^2$ invariant subspace under consideration
(compare also Ref.~[8]): 
\beq
(p^\nu\gamma_\nu -m) u_\mu (\p )&=&0\, , 
\label{Dirac_RS_2}\\
W^\mu u_\mu (\p )&=&0\, .
\label{PaLu_SC}
\eeq
The propagator of the Roper resonance resulting from Eqs.~(\ref{Dirac_RS_2}), (\ref{PaLu_SC}) 
would be
\beq
{\mathcal S}^{Roper}_{\mu\nu }=
{{( - {{p_\mu p_\nu}\over m^2})(p^\lambda\gamma_\lambda +m)}
\over {2m(p^2-m^2)}}\, .
\label{Rop_prop}
\eeq
Eventually, the Roper resonance could  be treated independently
from Eqs.~(\ref{Dirac_RS_2}),~(\ref{PaLu_SC}), and (\ref{Rop_prop}) 
by means of the Dirac equation alone.
 
A different  perspective on $N(1520)-N(1535)$ appears
in recalling that the sixteen dimensional vector representing the
direct product of the four-vector with the Dirac spinor,
$\Psi (\p )= A_\mu (\p ) \otimes \psi (\p ) $, is reducible into
\beq
\Psi (\p ): \,\, \left( {1\over 2}, {1\over 2} \right)\otimes
{\Big[} \left( {1\over 2},0 \right)\oplus \left( 0,{1\over 2}\right)
{\Big]}\longrightarrow
{\Big[} \left(1,{1\over 2}\right)\oplus \left({1\over 2},1\right){\Big]}
\oplus 
{\Big[} \left({1\over 2},0\right)\oplus \left(0,{1\over 2}\right){\Big]}\, .
\nonumber\\
\label{psimu_red}
\eeq
In this case, the Roper has clearly to be attached to the Dirac piece, 
while the $D_{13}-S_{11}$ sub-cluster has to be mapped onto, 
${\Big[}\left(1,{1\over 2}\right)\oplus \left({1\over 2},1\right){\Big]}$.
There are different ways to obtain the
wave equations for the latter representations (see Napsuciale's talk 
in Ref.~[12]). In one of the possibilities one may consider
the wave equation for $\Psi (\p ) $ as the direct product of the
wave equations for the four vector, presented in our previous work,$^8$ 
on the one side, and the Dirac equation, on the other,
\beq
\left(\Lambda_{\mu\nu}p^\mu p^\nu  \pm  m^2 1_4\right)\otimes
\left( \gamma^\eta p_\eta \mp m 1_4 \right)\Psi (\p ) &=& 0\, ,
\label{Psi16_eq}
\eeq
with the matrices $\Lambda_{\mu\nu}$ from Ref.~[8].
Now, by means of an appropriate similarity transformation, the 
16$\times $16 dimensional matrix in front of $\Psi (\p )$ can be 
block-diagonalized as to obtain a 12$\times $12 matrix for
${\Big[}\left(1,{1\over 2}\right)\oplus \left({1\over 2},1\right){\Big]}$,
(and thereby the corresponding wave equation)
on top of Dirac's 4$\times$4 matrix $(p^\mu\gamma_\mu \mp m 1_4)$ for
$ ({1\over 2},0)\oplus (0,{1\over 2})$.  The disadvantage of this scheme
lies in the loss of separation between space-time--  and spinor indices,
on the one side, and in shoveling problems of indefinite metrics onto
the 12 component rest space, on the other. 
The subject is currently under investigation.
Nonetheless, by means of Eqs.~(\ref{Dirac_RS_1}),~(\ref{Proca_SC}), 
and (\ref{PaLu_SC}) 
all the  $\widetilde{W}^2$-- invariant subspaces 
of $\psi_\mu (\p )$ (including the parity degenerate one) are described as 
physical  and the idea of $\psi_\mu (\p )$ (be it $u_\mu (\p )$, or,
$v_\mu (\p )$) as a particle superior
of states characterized by unique $SU(2)$ spins, and unique 
fermion numbers, still preserves viability.\\
   
The present lecture devotes itself to reviewing theory and phenomenology
of $\psi_{\mu_1...\mu_K}(\p )$  states.
The presentation is organized as follows. In the next Section
I present existing observations on baryon spectra.
In Section III I compare different
ideas on data interpretation. There I also review
the model of a rotating and vibrating quark-diquark system$^{13}$ 
and the agreement of its predictions with  observations.
Also there, in following Refs.~[14]-- [17],
I review QCD inspired motivations for legitimacy
of a quark-diquark configuration in baryon structure.
Section IV is devoted to the auxiliary condition
$\gamma^\mu\psi_\mu (\p )$ of the Rarita-Schwinger framework$^{18}$
for the four-vector--spinor. There I show that $\gamma^\mu\psi_\mu (\p )=0$ 
is a short-hand from the general definition of the parity singlet
$\widetilde{W}^2$-- invariant subspace $\psi_\mu^{\tau^+_1} (\p )$
by means of the covariant projector, 
$-{1\over 3}({1\over m^2}\widetilde{W}^2 +{3\over 4})\psi_\mu^{\tau^+_1}(\p )=
\psi_\mu^{\tau^ +_1} (\p )$. 
I suggest solving the latter equation for $\psi_\mu^{\tau^+_1}(\p )$,
which is of second order in the momenta and does not need any further
supplementary conditions (provided,  $\psi_\mu (\p )$ is considered
as a 16 dimensional vector) rather than using the Rarita-Schwinger 
set of linear differential equations 
$(p^\nu\gamma_\nu -m)\psi_\mu (\p )=0$,
and $p^\mu\psi_\mu (\p )=0$, as supplemented by
$\gamma^\mu\psi_\mu (\p ) =0$. In so doing, one 
derives in Section IV the associated  Lagrangian 
and verifies that our scheme does not suffer the
Velo-Zwanziger$^{19}$ problem of complex energy in the presence of an 
electromagnetic field
and acausal propagation.
The paper ends with a brief outlook.

\section{Spectra of light-quark baryons}
\subsection{Observations}

Understanding the spectrum of the most simplest composite systems has always
been a key point in the theories of the micro-world. Recall that quantum
mechanics was established only after the successful description of the 
experimentally observed regularity patterns (such like the Balmer- series)
in the excitations of the hydrogen atom. Also in solid state physics, the 
structure of the low--lying excitations, be them without or with a gap,
has been decisive for unveiling the dynamical properties of
the many-body system-- ferromagnet versus superconductor, and the
relevant degrees of freedom, magnons versus Cooper pairs.

In a similar way, the regularity patterns of the nucleon excitations
are decisive for uncovering the relevant subnucleonic degrees of freedom 
and the dynamical properties of the theory of strong interaction-- 
the Quantum Chromo- Dynamics.

Despite its long history, amazingly, the structure of the nucleon spectrum
is far from being settled. This is due to the fact that the first facility
that measured nucleon levels, the Los Alamos Meson Physics Facility 
(LAMPF) did not find all the states that were predicted by the excitations 
of three quarks. Later on, the Thomas Jefferson National Accelerator Facility
(TJNAF) was designed (among others) to search for those 
``missing resonances''.
At present, all data have been collected and are awaiting evaluation.$^{21}$  

In a series of papers$^{6}$ a new and subversive look on the
reported data in Ref.~[22] was undertaken. There I drew attention 
to the ``{\tt Come-Together}'' of resonances of different spins 
and  parities to narrow mass bands in the nucleon spectrum and, its exact
replica in the  $\Delta $ spectrum (see Figs.~ 1 and 2). 

The first group of states consists of two spin-${1\over 2}$
states of opposite parities and a parity singlet spin-${3\over 2}^-$.
The second group has three parity degenerate states with spins
varying from ${1\over 2}^\pm$ to ${5\over 2}^\pm $, and a 
single spin-${7\over 2}^+$ state.
Finally, the third group has five parity degenerate states with spins ranging
from ${1\over 2}^\pm$ to ${9\over 2}^\pm $, and a single spin 
${{11}\over 2}^+$ state 
(see Ref.~[23] for the complete $N$ and $\Delta $ spectra).
A comparison between the $N$ and $\Delta $ spectra shows that they
are identical up to two ``missing'' resonances
on the nucleon side (these are the counterparts of the $F_{37}$ and $H_{3, 11}$
states of the $\Delta $ excitations) and up to three ``missing'' states on 
the $\Delta $ side (these are the counterparts of the nucleon 
$P_{11}$, $P_{13}$, and $D_{13}$ states from the third group). 
The $\Delta (1600)$ resonance
which is most probably and independent hybrid state, is the only state
that at present seems to drop out of our systematics.

\subsection{Ideas}

The existence of exactly same nucleon- and $\Delta $ crops of resonances 
raises several questions:

\begin{enumerate}
\item  Are we facing here a new type of symmetry which was not anticipated
       by any model or theory before?
\item  Is the clustering pattern a pure accident?
\item Or, is it an artifact of spectra incompleteness?
\end{enumerate}

The oldest idea favors the possibility of spectrum incompleteness
and counts on the discovery of ``missing'' resonances that are supposed to
restore the uniform distributions of excitations in the spectra
in accordance with the predictions of the three-quark model.  
A more recent idea puts the emphasis on the tendency of resonances
to  pair and interpret the pairing as a signal for 
manifest chiral symmetry.

Indeed, it is hardly to overlook the existence of parity couples
\begin{eqnarray*}
N\left({1\over 2}^+;1440\right) \mbox{--} 
N\left({1\over 2}^-;1535\right)\, , &\quad &   
\Delta\left({1\over 2}^+;1900\right) \mbox{--} 
\Delta\left({1\over 2}^-;1910\right)\, ,\\ 
N\left({3\over 2}^+;1720\right) \mbox{--} 
N\left({3\over 2}^-;1700\right)\, , &\quad &  
\Delta\left( {3\over 2}^+;1920\right) \mbox{--}
\Delta\left( {3\over 2}^-;1940\right)\, \\
N\left(({5\over 2}^+;1675\right) \mbox{--} 
N\left( {5\over 2}^-;1680\right)\, , &\quad& 
\Delta\left( {5\over 2}^+;1905\right) \mbox{--} 
\Delta\left( {5\over 2}^-;1930\right)\,, \\
N\left({9\over 2}^+;2200\right) \mbox{--} 
N\left( {9\over 2}^-;2250\right)\, , &\quad&   
\Delta\left({9\over 2}^+;2300\right) \mbox{--}
\Delta\left({9\over 2}^-;2400\right)\, ,
\end{eqnarray*}
and escape the temptation to interpret them as parity doublets.
The tendency of resonances to form  parity couples was realized
in the very early days of nucleon spectroscopy. It would be a long
list of literature would one want to compile a complete bibliography
on that subject. I only like to mention at this place work by Dashen,$^{24}$ 
D\"onau and Reinhardt,$^{25}$, 
Iachello,$^{26}$, and Robson.$^{27}$
Yet,  new is often what has been only well forgotten
and eagerly awaiting for a remake. In 1992 
I was still under the very strong impression, the 1990 Nobel 
prize to E. J. Corey for his achievements on the synthesis of chiral 
drugs had on me, and very much enthusiastic about chirality in biology. 
The chiral symmetry of laboratory manufactured chemical 
(better, stereo-chemical) substances was the father of my wish
to find their counterpart in the particle world and what could be better
suited for that but the resonant parity couples.   
In a 1992 preprint Ref.~[28] I wrote:
\begin{quote}
``{\it In studying chiral symmetry of strong interacting particle systems 
one can not oversee their analogy to enantiomorphic inorganic systems 
considered in stereo-chemistry. The approximate duplication of the nucleon 
excitations with respect to parity observed above 1.3 GeV can be viewed 
as a kind of enantiomorphism at the resonance particle level. \/}''
\end{quote}
In my lectures$^{29}$ on chiral symmetry
at the Indian Summer School 1992 in Sassava, former Czechoslovakia
I restricted myself to the more factual
expression for same reality by saying that
 \begin{quote}
``{\it Chiral symmetry is indeed realized (at least partly) in the excitation
spectrum of the non-strange baryons with masses higher but 1.3 GeV, 
where the duplication of the resonances with respect to parity is 
well pronounced\/}''.
\end{quote}
The parity doublet ``sound-track'' was taken on in good spirit
by D. O. Riska and found extensive coverage 
in his lectures at the Swieca school on nuclear
physics in Brazil$^{30}$ early 1993.  
The years that followed are a symptomatic example for 
a hasty and uncritical remake of an idea by various groups, 
and its subsequent promotion to the ultimate truth,
for the good or bad.$^{31-35}$
Although questioning seems to be out of question in the scene,
the unprejudiced reader is invited to have his/her own
independent look onto the following questions:
\begin{enumerate}
\item What drives parity couples of {\it different spins\/}
to share same narrow mass bands?
\item Can manifest chiral symmetry be realized   ``{\it partly}'' ,
or ``{\it approximately\/}'' ?
Stated differently,  does the systematic occurrence
 of  ``{\it unpaired}'' states like
$N \left({3\over 2}^-;1520\right)$, 
$\Delta \left({7\over 2}^+;1950\right)$, 
$\Delta \left({3\over 2}^-;1650\right)$,
$\Delta \left( {{11}\over 2}^+;2420\right)$ destroy chiral symmetry?
\item In Ref.~[27], a comment on Iachello's paper Ref.~[26] 
on parity doublets in hadron spectra, Robson warns that the
separation between angular momentum and intrinsic spin performed
in the contemporary quark models, is incompatible with relativity.
\end{enumerate}
The subsequent Section attends to these questions.

\section{Baryon systematics in terms of $SU(2)\otimes O(4)$ multiplets}
As already mentioned in Section I, the excitations of the
light-quark baryons fit exactly into multiplets of the type
$\left({K\over 2},{K\over 2}\right)\otimes 
{\Big[}\left( {1\over 2},0\right)\oplus \left(0,{1\over 2}\right){\Big]} $.
There are two possible avenues in treating such multiplets.
The first leads over the internal quark baryon dynamics and ends up
in looking upon such states as composite systems 
characterized by a  four dimensional angular momentum $K$.  
The second goes over relativistic states transforming co-variantly 
from one inertial frame to an other as totally symmetric Lorentz tensors
of rank-$K$ with Dirac spinor components.
This is akin to the modeling of the nucleon structure
when one first couples the three spin-${1\over 2}$ 
constituent quarks to a spin-${1\over 2}$  nucleon, and then 
considers the nucleon as covariant  
$\left( {1\over 2},0\right) \oplus \left( 0,{1\over 2}\right)$ 
Lorentz state (up to form factors).

The present Section is devoted to the first avenue while the next
Section deals with the second option.

\subsection{The quark version of the diatomic rovibron model and the
clustering in baryon spectra}
 
Baryons in the quark model are considered as constituted of three
quarks in a color singlet state. It appears naturally, therefore,
to undertake an attempt of describing the baryonic system
by means of algebraic models developed for
the purposes of triatomic molecules,
a path already pursued by  Refs.~[36].
There, the three body system was described in terms of two
vectorial ($\vec{p}\, ^+$) and one scalar $(s^+)$ boson degrees of freedom
that transform as the fundamental  $U(7)$ septet.
In the  dynamical symmetry limit 
\begin{equation}
U(7)\longrightarrow U(3)\times U(4)\, ,
\label{U(7)}
\end{equation}
the degrees of freedom associated with the one vectorial boson
factorize from  those associated with
the scalar boson and the remaining vectorial boson.
Because of that the physical states constructed within the $U(7)$ IBM model
are often labeled by means of  $U(3)\times U(4)$ quantum numbers.
Below we will focus on that very sub-model of the IBM and
show that its excited modes exactly  accommodate the K--clusters 
from above and thereby the LAMPF data on the non-strange baryon
resonances.  

The dynamical limit $U(7)\longrightarrow U(3)\times U(4)$ corresponds
to the quark--diquark approximation$^{37}$ of the three quark system, 
when two of the quarks reveal a stronger pair correlation to a 
diquark (Dq), while the third quark (q) acts as a spectator.
The diquark approximation turned out to be rather convenient
in particular in describing various properties of the ground state 
baryons.$^{38-39}$ Within the context of the quark--diquark (q-Dq) 
model, the ideas of the rovibron model, known from the spectroscopy of 
diatomic molecules,$^{40}$ can be applied to the description
of the rotational-vibrational (rovibron) excitations of the q--Dq system.

\subsection{Rovibron Model for the Quark--Diquark System}

In the rovibron model (RVM) the relative q--Dq motion 
is described by means of four types of boson creation
operators $s^+, p^+_1, p^+_0$, and $p^+_{-1}$ 
(compare$^{40}$). The operators $s^+$ and $p^+_m$ in turn 
transform as rank-0, and rank-1 spherical tensors,
i.e. the magnetic quantum number $m$ takes
in turn the values $m=1$, $0$, and $-1$.
In order to construct boson-annihilation operators that
also transform as spherical tensors, one introduces
the four operators $\widetilde{s}=s$, and
$\widetilde{p}_m=(-1)^m\, p_{-m}$.
Constructing rank-$k$ tensor product of
any rank-$k_1$ and rank-$k_2$ tensors, say, $A^{k_1}_{m_1}$ 
and $A^{k_2}_{m_2}$, is standard and given by
\begin{equation}
\lbrack A^{k_1}\otimes A^{k_2}\rbrack^k_m =
\sum_{m_1,m_2}(k_1m_1 k_2m_2\vert km)\, A^{k_1}_{m_1}A^{k_2}_{m_2}\, .
\label{clebsh}
\end{equation}
Here, $(k_1m_1k_2m_2\vert km)$ are the well known $O(3)$ Clebsch-Gordan
coefficients.

Now, the lowest states of the two-body system are identified with $N $
boson states and are characterized by the ket-vectors 
$\vert n_s\, n_p\, l\, m\rangle $ (or, a linear combination of them)
within a properly defined Fock space. The constant  
$N=n_s +n_p$ stands for the total number of $s$- and $p$ bosons
and plays the r\'ole  of a parameter of the theory.
In molecular physics, the parameter $N$ is usually associated
with the number of molecular bound states.
The group symmetry of the rovibron model is well known to be $U(4)$. 
The fifteen generators of the associated $su(4)$ algebra 
are determined as the following set of bilinears 
\begin{eqnarray}
A_{00}=s^+ \widetilde{s}\, , &\quad& A_{0m}=s^+ \widetilde{p}_m\, , 
\nonumber\\
A_{m0}=p^+_m\widetilde{s}\, , &\quad & A_{mm'}=p^+ _m
\widetilde{p}_{m'}\, .
\label{RVM_u4}
\end{eqnarray} 
The $u(4)$ algebra is then recovered by the following 
commutation relations
\begin{equation}
\lbrack A_{\alpha\beta},A_{\gamma\delta}\rbrack_-=
\delta_{\beta \gamma}A_{\alpha\delta}-
\delta_{\alpha\delta}A_{\gamma\beta}\, .
\end{equation}
The operators associated with physical observables can then be expressed
as combinations of the $u(4)$ generators.
To be specific, the three-dimensional angular momentum takes the form
\begin{equation}
L_m=\sqrt{2}\, \lbrack p^+ \otimes \widetilde{p}\rbrack^1_m \, .
\label{a_m}
\end{equation}
Further operators are  $(D_m$)-- and $(D'_m$) defined as 
\begin{eqnarray}
D_m &=&\lbrack p^+\otimes \widetilde{s}+s^+\otimes 
\widetilde{p}\rbrack^1_m\, ,\\
\label{x_dipol_rvm}
D_m '&=&i\lbrack p^+\otimes \widetilde{s}-s^+\otimes 
\widetilde{p}\rbrack^1_m\, ,
\label{p_dipol_rvm}
\end{eqnarray}
respectively.
Here, $\vec{D}\, $ plays the r\'ole of
the electric dipole operator.

{}Finally, a quadrupole operator $Q_m$ can be constructed as
\begin{equation}
Q_m=\lbrack p^+\otimes \widetilde{p}\rbrack^2_m\, ,
\quad \mbox{with}\quad m=-2,..., +2\, .
\label{quadr_rvm}
\end{equation}
The $u(4)$ algebra has the two algebras $su(3)$, and $so(4)$, as respective
sub-algebras. The $su(3)$ algebra is constituted by the three
generators $L_m$, and the five components of the quadrupole operator
$Q_m$. Its $so(4)$ sub-algebra is constituted by the 
three components of the angular momentum operator $L_m$, on the one side,
and the three components of the operator $D_m'\, $, on the other side. 
Thus there are two exactly soluble RVM limits that correspond 
to the two different chains of reducing
$U(4)$ down to $O(3)$. These are:
\begin{equation}
 U(4)\supset U(3)\supset O(3)\, ,\quad \mbox{and} \quad  
U(4)\supset O(4)\supset O(3)\, ,
\label{chains}
\end{equation}
respectively.
The Hamiltonian of the RVM in these exactly soluble limits is 
then constructed as a properly chosen function of the Casimir 
operators of the algebras of either the first, or the second chain. 
{}For example, in case one approaches $O(3)$ via $U(3)$,
the Hamiltonian of a dynamical $SU(3)$ symmetry can be cast into 
the form:
\begin{equation}
H_{su(3)} = H_0 +\alpha\,  {\cal C}_2\left( su(3) \right) +
\beta \, {\cal C}_2\left( so(3) \right)\, .
\label{H_SU3}
\end{equation}
Here, $H_0$ is a constant,
${\cal C}_2\left( su(3) \right)$, and ${\cal C}_2\left( so(3) \right)$ are 
in turn the quadratic (in terms of the generators) 
Casimirs of the $su(3)$, and $so(3)$ algebras, 
respectively, while $\alpha $ and $\beta $ are constants, to
be determined from data fits.

A similar expression (in obvious notations) can be written for the 
RVM Hamiltonian in the $U(4)\supset O(4)\supset O(3)$ exactly soluble limit:
\begin{equation}
H_{so(4)} = H_0 +\widetilde{\alpha}\, {\cal C}_2\left(so(4)\right) +
\widetilde{\beta} {\cal C}_2\left(so(3)\right)\, .
\label{H_SO4}
\end{equation}
The Casimir operator ${\cal C}_2\left(so(4)\right)$ is defined accordingly as
\begin{equation}
{\cal C}_2\left( so(4)\right)={1\over 4}\left( \vec{L}\, ^2 + 
\vec{D}\, ' \, ^2
\right)\, 
\label{so(4)_Casimir}
\end{equation}
and has an eigenvalue of ${K\over 2}\left( {K\over 2}+1 \right)$.
In molecular physics, only linear combinations of the Casimir operators 
are used, as a rule. However, as known
from the hydrogen atom,$^{41}$ the Hamiltonian is determined by the 
inverse  power of ${\cal C}_2\left(so(4)\right) $ according to
\begin{equation}
H_{Coul}=f \left( -4{\mathcal C}_2\left(so(4)\right) -1\right)^{-1}
\label{Coul+SO(4)}
\end{equation}
where $f$ is a parameter with the dimensionality of mass.
This Hamiltonian predicts the energy of the states as $E_K=-f/(K+1)^2$ 
and does not follow the simple linear pattern
(see also Eq.~(\ref{H_SO4})).

In order to demonstrate how the RVM applies to baryon spectroscopy,
let us consider the case of q-Dq states associated with $N=5$
and for the case of a $SO(4)$ dynamical symmetry. 
{}From now on we shall refer to the quark rovibron model
as qRVM. It is of common
knowledge that the totally symmetric irreps of the $u(4)$ algebra 
with the Young scheme $\lbrack N\rbrack$ contain the 
$SO(4)$ irreps $\left({K\over 2}, {K\over 2}\right)$ with
\begin{equation}
K=N, N-2, ..., 1 \quad \mbox{or}\quad 0\, .
\label{Sprung_K}
\end{equation}
Each one of these $SO(4)$ irreps contains $SO(3)$ multiplets with
three dimensional angular momenta
\begin{equation}
l=K, K-1, K-2, ..., 1, 0\, .
\label{O3_states}
\end{equation}
In applying the branching rules in Eqs.~(\ref{Sprung_K}),
(\ref{O3_states})
to the case $N=5$, one encounters the series of levels
\begin{eqnarray}
K&=&1: \quad l=0,1;\nonumber\\
K&=&3: \quad l=0,1,2,3;\nonumber\\
K&=&5: \quad l=0,1,2,3,4,5\, .
\label{Ns_Ks}
\end{eqnarray}
The parity carried by these levels is $\eta (-1)^{l}$ where
$\eta $ is the parity of the relevant vacuum. In coupling now the
angular momenta in Eq.~(\ref{Ns_Ks}) to the spin-${1\over 2}$ of the three
quarks in the nucleon, the following sequence of states is obtained:
\begin{eqnarray}
K&=&1: \quad \eta J^\pi={1\over 2}^+,{1\over 2}^-, {3\over 2}^-\, ;
\nonumber\\
K&=&3: 
\quad \eta J^\pi={1\over 2}^+,{1\over 2}^-, {3\over 2}^-,
{3\over 2}^+, {5\over 2}^+, {5\over 2}^-, {7\over 2}^- \, ;
\nonumber\\
K&=&5: \quad \eta J^\pi={1\over 2}^+,{1\over 2}^-, {3\over 2}^-,
{3\over 2}^+, {5\over 2}^+, {5\over 2}^-, {7\over 2}^- ,
{7\over 2}^+, {9\over 2}^-, {11\over 2}^-\, .
\label{set_q}
\end{eqnarray}
Therefore, rovibron states of half-integer spin transform according to  
$\left( {K\over 2},{K\over 2}\right) \otimes \left[
\left({1\over 2},0 \right) \oplus 
\left( 0,{1\over 2} \right)\right] $
representations of $SO(4)$.
The isospin structure is accounted for pragmatically through
attaching to the K--clusters an isospin spinor
$\chi^I$ with $I$ taking the values $I={1\over 2}$ 
and $I={3\over 2}$ for the nucleon, and the $\Delta $ states,
respectively.
As illustrated by Figs.~1 and 2, the above quantum numbers cover
both the nucleon and the $\Delta $ excitations.

Note that in the present simple version of the
rovibron model, the spin of the quark--diquark system is
$S={1\over 2}$, and the total spin $J$ takes the values $J=l\pm {1\over 2}$
in accordance with Eqs.~(\ref{Ns_Ks}) and (\ref{set_q}).
The strong relevance of {\it same \/} picture for both
the nucleon and the $\Delta (1232) $ spectra (in $\Delta (1232)$
the diquark is, however,  in a vector-isovector state) hints onto the
dominance of a scalar (pseudoscalar) diquark for both the 
excited nucleon-- and  $\Delta (1232)$ states.
This situation is reminiscent of the $^210$ configuration
of the $70(1^-)$plet of the canonical 
$SU(6)_{SF}\otimes O(3)_L$ symmetry
where the mixed symmetric/antisymmetric character
of the $S={1\over 2}$ wave function in spin-space is compensated
by a mixed symmetric/antisymmetric wave function in coordinate space,
while the isotriplet $I={3\over 2}$ part is totally symmetric.

We here will leave aside the discussion of the generic problem
of the various incarnations of the IBM model regarding the
symmetry properties of the resonance wave functions to a later
date and rather concentrate in the next subsection onto 
the ``missing'' resonance problem.

\subsection{Observed and ``Missing'' Resonance Clusters within the 
Rovibron Model}

The comparison of the states in Eq.~(\ref{set_q}) with the reported ones
in  Figs.~1 and ~2 shows that the predicted sets reproduce exactly
the quantum numbers of the non-strange baryon excitations with
masses below $\sim $ 2500 MeV, provided, the parity $\eta $ of the vacuum
changes from scalar ($\eta =1$) for the $K=1$, to pseudoscalar 
($\eta =-1$) for the $K=3,5$ clusters.
A pseudoscalar ``vacuum'' can be modeled in terms of
an excited  composite  diquark carrying an internal 
angular momentum  $L=1^-$ and maximal spin $S=1$. In one of
the possibilities the total spin of such
a system can be $\vert L-S\vert = 0^-$.

To explain the properties of the ground state, one has to
consider separately even $N$ values, such as, say, $N'=4$.
In that case another branch of excitations, with $K=4$, $2$, and
$0$ will emerge. The $K=0$ value characterizes the 
ground state,  $K=2$ corresponds to
$\left( 1,1\right)\otimes 
\lbrack\left( {1\over 2},0\right)\oplus 
\left(0,{1\over 2}\right) \rbrack $, while  $K=4$ corresponds to
$\left( 2,2\right) \otimes\lbrack\left( {1\over 2},0\right)\oplus 
\left(0,{1\over 2}\right) \rbrack $. 
These are the multiplets that we 
will associate with the  ``missing'' resonances  
predicted by the rovibron model.
In this manner, reported and ``missing'' resonances  fall apart 
and populate  distinct $U(4)$- and $SO(4)$ representations.
In making observed and ``missing'' resonances distinguishable,
reasons for their absence or, presence in the spectra are
easier to be searched for.
As to the parity of the resonances with even $K$'s, there is some ambiguity.
As a guidance one may consider the decomposition of the three-quark
($q^3$) Hilbert space into Lorentz group representations
as performed in Ref.~[42]. There, two states of 
the type
$\left( 1,1\right)\otimes 
\lbrack\left( {1\over 2},0\right)\oplus 
\left(0,{1\over 2}\right) \rbrack $
were found. The first one arose out of the decomposition of 
the $q^3$-Hilbert space spanned by the $1s-1p-2s$ single-particle
states. It was close to  $\left( {1\over 2},{1\over 2}\right)\otimes 
\lbrack\left( {1\over 2},0\right)\oplus 
\left(0,{1\over 2}\right) \rbrack $
and carried opposite parity to the latter.
It accommodated, therefore, 
unnatural parity resonances. 
The second $K=2$ state was part of the
$(1s-3s-2p-1d)$- single-particle configuration space and
was closer to $\left( {3\over 2},{3\over 2}\right)\otimes 
\lbrack\left( {1\over 2},0\right)\oplus 
\left(0,{1\over 2}\right) \rbrack $.
It also carried opposite parity to the latter and accommodated
natural parity resonances. Finally, the $K=4$ cluster 
$\left( 2,2\right)\otimes 
\lbrack\left( {1\over 2},0\right)\oplus 
\left(0,{1\over 2}\right) \rbrack $
emerged in the decomposition of the one-particle-one-hole states within
the $(1s-4s-3p-2d-1f-1g)$ configuration space and
carried also natural parity, that is, opposite parity to
  $\left( {5\over 2},{5\over 2}\right)\otimes 
\lbrack\left( {1\over 2},0\right)\oplus 
\left(0,{1\over 2}\right) \rbrack $.
In accordance with the above results,
we here will treat the $N=4$ states to be all of natural parities
and identify them with the nucleon $(K=0)$, the natural parity $K=2$,
and  the natural parity $K=4$--clusters.

The unnatural parity $K=2$ cluster from$^{42}$
could be generated through an unnatural
parity $N=2$ excitation mode. 
However, this mode would require manifest chiral symmetry up to 
$\approx 1550$
MeV which contradicts at least present data.
With this observation in mind, we here will restrict ourselves
to the consideration of the natural parity $N=4$ clusters.
In this manner the unnatural parity $K=2$ state from Ref.~[42]
will be dropped out from the current version of the rovibron model.
{}From now on we will refer to the excited $N=4$ states as to
``missing'' rovibron  clusters.

Now, the qRVM Hamiltonian that fits masses of the reported
cluster states is given by the following
function of ${\cal C}_2\left(so(4)\right)$
\begin{equation}
H_{qRVM}=H_0 - f_1\, \left(4 {\cal C}_2\left( so(4)\right) +1\right)^{-1}
+f_2\,  {\cal C}_2(so(4) \, .
\label{H_QRVM}
\end{equation}
The states in  Eq.~(\ref{set_q}) are degenerate and the dynamical symmetry
is $SO(4)$. Here, the parameter set has been chosen as  
\begin{equation}
H_0= M_{N/\Delta } +f_1\, ,\quad  f_1=600\, \, \mbox{MeV}\, , 
\quad f_2^N=70\, \, \mbox{MeV}\, ,\quad
f_2^\Delta= 40\, \, \mbox{MeV}\, .
\label{f_s}
\end{equation}
Thus, the $SO(4)$ dynamical symmetry limit of the
qRVM picture of baryon structure motivates  
existence of quasi-degenerate crops of resonances 
in both the nucleon- and $\Delta $ baryon spectra.
In Table I we list the masses of the K--clusters concluded from
Eqs.~(\ref{H_QRVM}), and (\ref{f_s}).
Preliminary indications for ``missing'' resonances have been reported
recently, for example,  in Refs.~[43], and [44].

The data on the $\Lambda $, $\Sigma $, and $\Omega^-$ hyperon spectra
are still far from being as complete as those of the nucleon and the 
$\Delta $ baryons and do not allow, at least at the present stage, 
a conclusive statement on relevance or irrelevance
of the rovibron picture.   
The presence of the heavier strange quark can significantly 
influence the excitation modes of the $q^3$-system.
In case, the presence of the $s$ quark in the
hyperon structure is essential,  the
$U(4)\supset U(3)\supset O(3)$ chain can be favored
over $U(4)\supset O(4)\supset O(3)$ and a different clustering motif
can appear here. For the time being, this issue
will be dropped out of the present consideration.

\subsection{Spin and quark--diquark in QCD}
The necessity for having a quark--diquark configuration within the
nucleon follows directly from QCD arguments. 
In Refs.~[14],~[15], ~[16] the notion of 
spin in QCD was re-visited in connection with the proton spin puzzle.
As it is well known, the spins of the valence quarks are by themselves not
sufficient to explain the spin-${1\over 2}$ of the nucleon.
Rather, one needs to account for
the orbital angular momentum of the quarks 
(here denoted by $L_{QCD}$) and the angular momentum carried
by the gluons (so called field angular momentum, $G_{QCD}$):
\begin{eqnarray*}
{1\over 2} &=&{1\over 2}\Delta \Sigma  +L_{QCD}  +G_{QCD} \, \\
= \int d^3 x {\Big[} {1\over 2}\bar \psi \vec{\gamma}\gamma_5\psi
&+&\psi^\dagger (\vec{x}\times (-\vec{D}))\psi 
+\vec{x}\times (\vec{E}^a\times \vec{B}^a){\Big]}\, .
\end{eqnarray*}
In so doing one encounters the problem that neither $L_{QCD}$, nor $G_{QCD}$
satisfy the spin $su(2)$ algebra. If at least  
$\left(L_{QCD} +G_{QCD}\right)$ is to do so,
\beq
{\Big[} 
\left( L_{QCD}^i +G^i_{QCD} \right),
\left( L_{QCD}^j +G^j_{QCD}\right) 
{\Big]}
=i
\epsilon^{ijk} \left( L_{QCD}^k +G^k_{QCD}\right)\, ,
\label{su2_alg}
\eeq	
then $E^{i;a}$ has to be  {\it restricted}  
to a chromo-electric charge, while $B^{i;a} $ 
has to be a chromo-magnetic dipole, 
\beq
E^{i;a}&=&{{gx' \,  ^i} \over {r'\, ^3}}T^a\, ,
\label{chromo_el}\\
B^{ia}&=& ({ {3x^i x^l m^l}\over r^5 } -
{m^i\over r^3})T^a\, ,
\label{chromo_mag}
\eeq
where  $x'\, ^i=x^i-R^i$.
In Singleton's contribution to this meeting (see Ref.~[15])
one reads that the diquark gives the color Coulomb fields,
while the quark gives the color magnetic dipole field.
In terms of color and flavor degrees of freedom,
the nucleon wave function indeed has the required quark--diquark
form:
\begin{eqnarray}
\vert p_{\uparrow} \rangle&=&
{ \epsilon_{ijk} \over \sqrt{18} }\,
\lbrack u^+_{i \downarrow} d^+_{j\uparrow} -
u^+_{i\uparrow}d^+_{j\downarrow} \rbrack\,  u^+_{k\uparrow}\,
\vert 0\rangle\, .
\label{di_quark_col}
\end{eqnarray}
A similar situation appears when looking for covariant QCD solutions
in form of a membrane with the three open ends being associated with the 
valence quarks. When such a membrane stretches to a string, so that
a linear action (so called gonihedric string) can be used, one again
encounters that very $K$-cluster degeneracies in the excitations 
spectra of the baryons, this time as a part of an infinite tower of states.
The result was  reported by Savvidy in Ref.~[17].
Thus the covariant spin-description provides an independent argument
in favor of a dominant quark-diquark configuration in the structure
of the nucleon, while search for covariant  
resonant QCD solutions leads once again to infinite $K$ cluster 
towers. 
The quark-diquark internal structure of the baryon's ground states
is just the configuration, the excited mode of which is described
by the rovibron model and which is the source of
the $\left( {K\over 2},{K\over 2}\right)\otimes
\Big[  \left( {1\over 2},0\right)\oplus \left(0,{1\over 2}\right)
\Big] $ pattern.

\subsection{ Manifest chiral symmetry and baryon spectra: 
parity doublets versus $\psi_{\mu_1...\mu_K}(\p )$ states}

At that stage we are ready to answer the questions posed to the end
of Section II. 
\begin{enumerate}
\item Several parity couples of different spins come together to 
same narrow mass band because
baryon excitations are rotational-vibrational modes of an 
excited quark-diquark string, be the diquark scalar 
(in the respective observed $\psi_\mu (\p )$, and the ``missing'' 
$\psi_{\mu_1\mu_2}(\p )$, and $\psi_{\mu_1...\mu_4} (\p )$), or 
pseudoscalar (in the observed $\psi_{\mu_1...\mu_3} (\p )$, and 
$\psi_{\mu_1...\mu_5} (\p )$), 
respectively.
Each $K$ state consists
of $K$ parity couples and a single unpaired ``has--been''  
spin-$J=K+{1\over 2}$ at rest. 
The parity couples should not be confused with parity doublets.
The latter refer to states of equal spins, residing in distinct
Fock spaces built on top of opposite parity (scalar, and pseudoscalar)
vacua with  $\Delta l=0$ .
The parity degeneracy observed in the baryon spectra is an artifact
of the belonging of resonances to 
$\left({K\over 2},{K\over 2}\right)\otimes 
{\Big[}\left({1\over2},0\right)\oplus \left(0,{1\over 2}\right){\Big]}$,
in which case the opposite parities of equal spins originate
from underlying angular momenta differing by one unit, i.e. from
$\Delta l=1$. Within the $K$--cluster scheme,
the unpaired states are no longer
of an uncertain status, but necessary for the completeness of the 
classification scheme.
The states of equal spins but opposite parities do not form parity doublets 
because the underling internal angular momenta
differ by one unit (i.e. $\Delta l =1$), while parity doublets
require $\Delta l=0 $.

The above considerations show that a $K$-mode of an
excited  quark-diquark string (be the diquark a scalar, or, pseudoscalar)
represents an independent entity (particle?) in its own rights which 
deserves its own 
name. To me the different spin facets of the $K$--cluster pointing into
different ``parity directions'' as displayed in Fig.~3 look like
barbs. That's why I suggest to refer to the $K$-clusters
as {\it barbed\/} states to emphasize the aspect of alternating parity. 
Barbs could also be associated with thorns 
(Spanish, espino), and {\it espinons\/} could be
another sound name for K-clusters.

\item Chiral symmetry realization within the $K$-cluster scenario
means having coexisting scalar and pseudoscalar diquarks
(''vacua''), and consequently coexisting $K$-clusters  of both
natural and unnatural parities.   
Stated differently, if TJNAF is to supplement the unnatural
parity LAMPF ``espinons'' with K=3, and 5 by the natural parity $K$=2,4
ones, then we will have manifest mode of chiral symmetry
in the baryonic spectra.  
The total number of ordinary-spin states in our scenario
needs not be multiple of two, as it should be in case of parity doubling.

\item 
In response to Robson's warning I here emphasize isomorphism
between non-relativistic rovibron excitations and relativistic
Lorentz group representations of the type
\beq
\left( {K\over 2},{K\over 2} \right)\otimes
{\Big[}\left({1\over 2},0\right)\oplus\left(0,{1\over 2} 
\right){\Big]} \simeq F^K (q^2) \, \psi_{\mu_1...\mu_K} (\p )\, ,
\label{o4_so(1,3)_mapping}
\eeq   
where $F^K (q^2)$ stands for an appropriate form factor
(or, a set of such).
Mapping non-relativistic $O(4)$ levels onto covariant
$\psi_{\mu_1...\mu_K}(\p )$ objects is not more and not less
justified but mapping Eq.~(\ref{di_quark_col}) onto,
$F (q^2) \, \left( {1\over 2},0\right)\oplus \left(0,{1\over 2}\right)$, 
where again, we denote form factors generically by $F (q^2)$.

\end{enumerate}

\subsection{Electromagnetic de-excitation modes of rovibron states}
In Ref.~[13] we presented  
the four dimensional Racah algebra that allows to calculate 
transition probabilities for electromagnetic de-excitations of
the rovibron levels. The interested reader is invited to consult
the quoted article for details. Here I restrict myself to reporting
the following two results.
\begin{enumerate}
\item All resonances from a $K$- mode have same widths.
\item As compared to the natural parity $K=1$ states,
the electromagnetic de-excitations of the unnatural parity
$K=3$ and $K=5$ rovibron states appear strongly suppressed.
\end{enumerate}  
To illustrate our predictions I compiled in Table 2 below
data on experimentally observed total widths of resonances 
belonging to  $K=3$, and $K=5$.

Table 2 clearly shows that resonances belonging to same $K$-mode have same
widths. 
The suppression of the electromagnetic de--excitation modes of unnatural
parity states to the nucleon (of natural parity)
is shown in Table 3.

The suppression under discussion is due to
the vanishing overlap between the scalar diquark  in the latter case, 
and a pseudo-scalar one, in the former. Non-vanishing widths can signal
small admixtures from natural parity states of same spins
belonging to even $K$ number states from the ``missing'' resonances.
For example, the significant value of $A_{{3\over 2}}^p$ for 
$N\left( {5\over 2}^+;1680\right)$
from $K=3$ may appear as an effect of mixing with the 
N$\left( {5\over 2}^+;1612\right)$ state
from the natural parity ``missing'' cluster with $K=2$,
\beq
|J^\pi={5\over 2}^+;m_{{5\over 2}}\rangle&=&
\sqrt{1-\alpha^2}|N=5;K=3; 0^-;l=2^-;{5\over 2}^+\, , m_{{5\over 2}}\rangle
\nonumber\\
&+&\alpha |N'=4;K=2; 0^+;l=2^+;{5\over 2}^+\, , m_{{5\over 2}}\rangle\, .
\label{mixing}
\eeq 
In first approximation, the mixing amplitude $\alpha $ that determines the
transition matrix element
\beq
\alpha\,  \langle N'=4;K=2;0^+;l=2^+;{5\over 2}^+\, , m_{{5\over 2}}|
T^{(1,1) 2m}| \mbox{gst}[N'=4;K=0;0^+;l=0^+];{1\over 2}^+\, , 
m_{{1\over 2}}\rangle\, ,
\label{tr_mae}
\eeq
between the $K$ mixed state and the nucleon can be considered
to be same for all resonances belonging to the cluster
under consideration (in the notations of  Ref.~[13]). 

This gives one the idea to
use helicity amplitudes to extract ``missing'' states.

\section{Pathology-free Lagrangians for $W^2$ invariant subspaces}

In this Section I focus onto the 
direct product space between the Lorentz vector,
$\left({1\over 2},{1\over 2}\right)$, 
generically denoted by $A_\mu (\p )$, and a Dirac spinor, 
$\left({1\over 2},0\right)\oplus \left(0,{1\over 2} \right)$:
\beq
\psi_\mu (\p ) =\left({1\over 2},{1\over 2}\right)\otimes 
\Big[\left({1\over 2},0\right)\oplus \Big[
\left(0,{1\over 2}\right) \Big]
:=A_\mu (\p ) \otimes \psi (\p )\, .
\label{vec_sp}
\eeq
Apparently,  for charged particles,
$\psi_\mu$ satisfies the Dirac equation
for any space-time index
\beq
(p^\nu\gamma_\nu -m)\psi_\mu (\p )=0\, .
\label{vsp_Dir}
\eeq
Spin- and parity content of $\psi_\mu (\p ) $ at rest is given by
\beq
\psi_\mu (\p )\, : \left( {1\over 2}^+; {1\over 2}^- , {3\over 2}^- \right)\, ,
\label{spin_par}
\eeq
where the example refers to natural  parity.
It is one of the long standing dreams of theoreticians to have
wave equations for particles with--higher spins that satisfy
the following four criteria:
\begin{enumerate}
\item The wave equations  are of pure arbitrary spin.
\item The wave equations  are linear in the momenta.
\item The wave equations allow for the direct
construction of vertices between the higher--spin state on the one 
side, and the nucleon--photon/pion system, on the other,
a demand best realized in terms of a separation between 
Lorentz (i.e. space-time) and Dirac (i.e. spinorial) indices. 

\item The wave equations do not suffer pathologies like having 
only non-standard energy-momentum dispersion relations and propagators.
\end{enumerate}
The Rarita-Schwinger framework in Eqs.~(\ref{Dirac_RS_1}), (\ref{Proca_SC}),
and (\ref{PaLu_SC}) satisfies the second  and third  criteria and violates
the first and fourth one. 

In the following I shall present an alternative to the 
Rarita-Schwinger framework that satisfies the third and fourth criteria,
and does not satisfy the first and second ones.
It has been shown by Ahluwalia {\it et al.\/} in Refs.~[45] 
that pure spin $(s,0)\oplus (0,s)$ states necessarily require wave equations
that are of the order $p^{2s}$ in the momenta. With that observation
in mind, the first criterion becomes irrelevant as it can not be fulfilled 
at all.
The avenue toward our goal leads over covariant projectors onto
invariant subspaces of the squared Pauli-Lubanski vector. In the following
subsections I shall first calculate the Pauli-Lubanski vector in 
$\psi_\mu (\p )$,
and then its squared. Afterward I shall present
the  construction of  covariant projectors
onto the $W^2$ invariant subspaces and emphasize that the 
$\gamma^\mu\psi_\mu (\p )=0$
subsidiary condition of the Rarita-Schwinger framework is a short-hand from
the more general definition of the unique $W^2$ invariant
subspace that is a parity singlet and of highest spin at rest.
Using that very definition as the principal
equation will turn out to be quite favorable. 
In so doing one ends up with a wave equation that is 
quadratic in the momenta (for $\psi_\mu (\p ) $ treated as a 16 dimensional
column vector, $\Psi (\p )$) and has a correct energy-momentum 
dispersion relation in the presence of a magnetic field thus 
avoiding the classical Velo-Zwanziger problem.

\subsection{The Pauli-Lubanski vector in $\psi_\mu (\p ) $}
In order to construct the Pauli-Lubanski vector for 
$A_\mu (\p )\otimes \psi (\p )  $ we first write down the generators
in $\left({1\over 2},0\right)\oplus \left(0,{1\over 2}\right)$  
\beq
S_{\nu\rho }^{\left({1\over 2},0\right)\oplus \left(0,{1\over 2}\right) }  =
{1\over 2}\sigma_{\nu\rho }\, ,&\quad&
\sigma_{\nu\rho}={i\over 2}\lbrack \gamma_\nu, \gamma_\rho \rbrack\, ,  
\label{Dirsmn}
\eeq
where $\gamma_\mu $ are the Dirac matrices.
The Lorentz generators in the four-vector 
$\left({1\over 2},{1\over 2}\right)$
space are obtained from those of the right-handed  
$\left( {1\over 2},0\right)$, and left-handed
$\left( 0, {1\over 2}\right)$ spinors$^{46}$
in noticing that $\left({1\over 2},{1\over 2}\right)$ is the direct product
of $\left( {1\over 2},0\right)$ and $\left( 0, {1\over 2}\right)$. 
With that in mind, one finds
\beq
S^{{1\over 2}{1\over 2}}_{0 l} = 
1_2 \otimes (i\sigma_l) +(- i\sigma_l)\otimes 1_2\, ,
\quad
S^{{1\over 2}{1\over 2}}_{ij}=
i\epsilon_{ijk}(1_2\otimes \sigma_k +\sigma_k\otimes 1_2)\, , 
\label{spinor_gen}
\eeq
where $\sigma_k$ are the standard Pauli matrices.
Now the generators in $\psi_\mu (\p ) $ are cast into the form 
(compare Ref.~[47] )
\beq
S_{\nu\rho} =S^{{1\over 2}{1\over 2}}_{\nu \rho}\otimes 1_4 +
1_4\otimes {1\over 2} \sigma_{\nu \rho}\, .
\label{psi_mu_gen}
\eeq

\subsection{The squared Pauli-Lubanski vector in $\psi_\mu (\p ) $}
Insertion of Eq.~(\ref{psi_mu_gen}) into Eq.~(\ref{PauLu}) 
shows that the Pauli-Lubanski vector in the product space
equals 
\beq
\widetilde{W}^
{ \left( {1\over 2}{1\over 2}\right)
\otimes 
\left[({1\over 2},0)\oplus(0,{1\over 2})\right]
}_\mu
=\widetilde{W}^{
{1\over 2}{1\over 2}
}_\mu
\otimes 1_4 + 1_4\otimes
W^{\left({1\over 2},0\right)\oplus\left(0,{1\over 2}\right)}
_\mu\, .
\label{w_mu_product}
\eeq
Apparently, $\widetilde{W}^2$, be they in
$\psi_\mu (\p )$, or the sixteen dimensional column vector, 
$\Psi (\p )$, split
these spaces into covariant sectors associated with
different $\widetilde{W}^2$ covariant eigenvalues according to 
\beq
{\Big(} 
\widetilde{W}^{{1\over 2}{1\over 2}}_\mu \otimes 1_4 +
1_4\otimes W^{ ({1\over 2},0)\oplus (0,{1\over 2})}_\mu
{\Big)}^2  \Psi^{\tau^\pm_l}(\p )  &=&
-\tau^\pm _l\,  m^2\,\Psi^{\tau^\pm_l} (\p )\, ,
\label{inv_subsp}
\eeq
where, again, $ \tau^\pm_l=(l\pm {1\over 2})(l\pm {1\over 2}+1)$,
and $l=K,K-1,...,1,0$. The parity non-degenerate 
$\tau^+_1=15/4$  sector describes a boosted ``{\it has--been\/}'' 
spin-${3\over 2}^-$ state in  the rest frame
and a state of indetermined spin elsewhere.
Equation ({\ref{inv_subsp}) is equivalently rewritten to
\beq
{\Big(}
\left(\widetilde{W}^{{1\over 2}{1\over 2}}\otimes 1_4\right)^2
+ \left(1_4\otimes W^{ ({1\over 2},0)\oplus (0,{1\over 2})}\right)^2
&+& 2 {\Big[}\widetilde{W}^{{1\over 2}{1\over 2}}\otimes 1_4{\Big]} 
\cdot
{\Big[}1_4\otimes W^{ ({1\over 2},0)\oplus (0,{1\over 2})}
{\Big]}{\Big)}\Psi^{\tau_l^\pm} (\p ) 
\nonumber\\
&=&\left(-\beta_l  m^2 -{3\over 4}m^2 -
\alpha^\pm_l m^2 \right)\Psi^{\tau_l^\pm} (\p )\nonumber\\
&=& -\tau^\pm_l m^2 \Psi^{\tau_l^\pm} (\p )\, ,
\label{a1}
\eeq
where $\beta_l$ is defined as
\beq
{\Big[}\widetilde{W}^{\left({1\over 2}{1\over 2}\right)}\otimes 1_4\,{\Big]}^2 
A  (\p )\otimes \psi(\p) =-\beta_l m^2\, A (\p )\otimes \psi(\p)\, ,
\label{beta_l}
\eeq
while
\beq
{\Big[}W^{\left({1\over 2},0\right)\oplus 
\left(0, {1\over 2}\right)}\otimes 1_4\,{\Big]}^2 
A (\p )\otimes \psi(\p) =-{3\over 4} m^2\, 
A  (\p )\otimes \psi(\p)\, .
\label{3_4}
\eeq
Finally,
\beq
2{\Big[}
 \widetilde{W}^{{1\over 2}{1\over 2}}\otimes 1_4{\Big]} \cdot
{\Big[}1_4\otimes W^{ ({1\over 2},0)\oplus (0,{1\over 2})}{\Big]}\, 
\Psi^{\tau_l^\pm} (\p )=-\alpha_l^\pm \, m^2 \Psi^{\tau_l^\pm} (\p )
\label{alpha_l}
\eeq
Obviously, equation (\ref{a1}) is satisfied if
\beq
\alpha_l^\pm =\tau_l^\pm -\beta_l -{3\over 4}\, .
\label{alpha_l_Gl}
\eeq
At this stage is important to recall that 
for any $\mu$, the component
$W^{{1\over 2}{1\over 2}}_\mu$ is represented by a $4\times 4$
matrix whose elements are again labeled by (two) Lorentz indices,
$\left(W^{{1\over 2}{1\over 2}}_\mu\right)_{\nu\eta}$,
as the components of the four dimensional $\left({1\over2 }{1\over 2}\right)$ 
vectors are labeled by a Lorentz index, i.e. $A_\mu (\p )$.

Insertion of the explicit expression for 
${\widetilde W}^{(1/2,0)\oplus(0,1/2)} $ into (\ref{w_mu_product}) 
and usage of the Dirac equation amounts to
\begin{eqnarray}
2{\Big[}
{\widetilde W}_\nu^{{1\over 2}{1\over 2}}\otimes 1_4{\Big]}
\, {\Big[} 1_4\otimes \left( -{1\over 4} \gamma_5 
\lbrack p\!\!\!/ ,  \gamma^\nu \rbrack_- \right){\Big]}
\Psi^{\tau_l^\pm}  (\p )   &=& -m^2 \alpha^\pm_l  \Psi^{\tau^\pm_l}  (\p), 
\nonumber\\
{\Big[} {\widetilde W}_\nu^{{1\over 2}{1\over 2}}\otimes 1_4{\Big]}
{\Big[} 1_4\otimes \gamma_5\left(p\!\!\!/\gamma^\nu -
\gamma^\nu p\!\!\!/\right){\Big]} \Psi^{\tau_l^\pm}  (\p)
 &=& 2m^2\alpha^\pm_l \Psi^{\tau^\pm_l} (\p\, ), \nonumber\\
{\Big[} {\widetilde W}_\nu^{{1\over 2}{1\over 2}}\otimes 1_4{\Big]}
\, {\Big[} 1_4\otimes \gamma_5
\left( 2g^{\alpha\nu}p_\lambda -2\gamma^\nu p\!\!\!/\right){\Big]}
\Psi^{\tau^\pm_l} (\p)  &=&2m^2\alpha^\pm_l  \Psi^{\tau^\pm_l} (\p),
\nonumber\\
{\Big[} {\widetilde W}^{{1\over 2}{1\over 2}}\cdot p
\, \otimes 1_4{\Big]}\, {\Big[} 1_4\otimes \gamma_5{\Big]}\,
 \Psi^{\tau_l^\pm} (\p) +
{\Big[}{\widetilde W}^{{1\over 2}{1\over 2}}\otimes 1_4{\Big]}\cdot
{\Big[} 1_4\otimes \gamma
\gamma_5  {\Big]}\,m\,  \Psi^{\tau_l^\pm} (\p) &=& 
m^2\alpha_l^\pm 
\Psi^{\tau_l^\pm} (\p ),
\nonumber\\
{\Big[} {\widetilde  W}^{{1\over 2}{1\over 2}}\otimes 1_4{\Big]}
\cdot  {\Big[} 1_4\otimes \gamma \gamma_5 {\Big]} \,\,  
\Psi^{\tau_l^\pm} (\p) &=& m 
\alpha^\pm_l \Psi^{\tau^\pm_l} (\p),
\label{Step1}
\end{eqnarray} 
where we used $\widetilde{W}^{{1\over 2}{1\over 2}}\cdot p=0$.

In going back to standard $\psi_\mu^{\tau^\pm_l}(\p )$ notation,
the last equation takes the form
\beq
{\Big[} {\widetilde  W}^{{1\over 2}{1\over 2}}_{\mu\eta }\otimes 1_4{\Big]}
\cdot  {\Big[} 1_4\otimes \gamma \gamma_5 {\Big]} \,\,  
\psi^{\tau_l^\pm}\, ^\eta (\p) &=& m 
\alpha^\pm_l \psi^{\tau^\pm_l}_\mu (\p)\, .
\label{po_komp}
\eeq
In taking now Lorentz contraction of both 
sides of Eq.~(\ref{po_komp}) by $\gamma^\mu$, one encounters the following
{\it constraint\/} for the $\gamma\cdot \psi^{\tau^\pm_l} (\p) $ spinor
\begin{eqnarray}
{1\over {m\alpha_l^\pm}}\gamma^\mu 
\left( {\Big[}
{\widetilde W}^{{1\over 2}{1\over 2}}_{\mu \eta }\otimes 1_4{\Big]}
\cdot 
{\Big[}\gamma\gamma_5 \otimes 1_4{\Big]}\right)
 \psi^{\tau_l^\pm}\,  ^\eta (\p) &=&  
\gamma\cdot \psi^{\tau^\pm_l} (\p) \, .
\label{Step2}
\end{eqnarray}
The non-relativistic counterpart of Eq.~(\ref{Step2}) reads
\begin{equation}
(1+{2\over \alpha^\pm_1 })\s \cdot  \vec{ \psi}^{\tau^\pm_1} (\0)  
=0\, .
\label{einblick}
\end{equation}
{}For $\alpha^+_1=1$ (related to $\tau^+_1=15/4$) one finds 
$\s\cdot \vec{ \psi}^{{{15}\over 4}}(\0)=0$ 
(corresponding to $\gamma\cdot \psi^{{ {15}\over 4}} (\p )=0$)
while for $\alpha^-_1=-2$  (related to $\tau^-_1=3/4$),
where the numerical factor in
(\ref{einblick}) vanishes, one encounters
$\s\cdot \vec{ \psi}^{{3\over 4}}(\0) \not=0$ 
(corresponding to $\gamma\cdot \psi^{{3\over 4}} (\p )\not= 0$).

Equation (\ref{Step2}) shows that the second auxiliary condition of
the Rarita-Schwinger framework is a short-hand of the 
general covariant definition of the single-parity has-been 
spin-${3\over 2}$ of $\psi_\mu (\p )$ as an invariant
subspace of the squared Pauli-Lubanski vector given in
Eq.~(\ref{w_mu_product}). 
In the subsequent Section I will exploit this fact in order
to show that the set of three equations in the Rarita-Schwinger
framework for a has--been spin-3/2 particle is favorably
replaced by a second order equation. 

\vskip 1cm


\subsection{Projectors onto $\widetilde{W}^2$ invariant subspaces}

On mass shell,  $p^2= m^2$, equations of the type of 
Eq.~(\ref{inv_subsp}) are equivalently cast into the form
(compare Ref.~[47]) 
\begin{eqnarray}
P^{{{15}\over 4}}   (\p) \Psi^{{{15}\over 4}}   (\p) &= &
\Psi  ^{{{15}\over 4}} (\p) \, ,\nonumber\\
P^{{{15}\over 4}}   (\p)  &= &-{1\over {3 }}\left[ 
{1\over m^2} \widetilde{W}^2+{3\over 4}  
\left( 1_4\otimes 1_4\right) \right]\, ,
\label{projs1}
\eeq
and
\beq
P^{{3\over 4}}  (\p) \Psi^{{3\over 4}}   (\p) &= &
\Psi  ^{{3\over 4}} (\p) \, ,\nonumber\\
P^{{3\over 4}}   (\p)  &= & +{1\over {3 }}\left[ 
{1\over m^2} \widetilde{W}^2 +{ {15}\over 4} 
\left( 1_4\otimes 1_4 \right) \right]\, .
\label{projs2}
\end{eqnarray}
In favor of transparent notations,
I here suppress the upper label
$\left( {1\over 2},{1\over 2}\right)\otimes
{\Big[}\left({1\over 2},0\right)\oplus \left(0,{1\over 2}\right){\Big]}$
in the squared Pauli-Lubanski vector in $A (\p )\otimes \psi (\p )$.

The non-relativistic version of these projectors has found application
in large $N_c$ baryon physics.$^{48}$ 
It is easily verified that the operators $P^{{{15}\over 4}}(\p ) $, and
$P^{{3\over 4}}(\p ) $ are indeed
projectors onto the $\widetilde{W}^2$ invariant 
subspaces with $\tau^+_1 ={{15}\over 4} $ 
(corresponding to $(1+{1\over 2})(1+{1\over 2}+1)$\, ),
on the one side, and onto $\tau^-_1={3\over 4}$ 
(corresponding to $(1-{1\over 2})(1-{1\over 2}+1)$\, ), 
and $\tau^\pm_0={3\over 4}$ (corresponding to 
$(0\pm {1\over 2})(0 \pm {1\over 2}+1)$\, ) respectively, 
on the other side, i.e.
\beq
{\Big[} P^{{{15}\over 4}} (\p){\Big]} ^2 = 
P^ {{{15}\over 4}} (\p) \, , &\quad& 
{\Big[} P^ {{{3}\over 4}}  (\p){\Big]}  ^2 = P^ {{{3}\over 4}}  (\p) \, , 
\nonumber\\
\quad P^{{{15}\over 4}} (\p)  +  P^{{{3}\over 4}} (\p) =1_{16}  \, ,
&\quad& P^{{{15}\over 4}} (\p)\, P^{{{3}\over 4}} (\p) =0\, .
\label{FNProj}
\eeq
Now, in making use of Eq.~(\ref{w2}), where we replaced the operators
$P^\mu$ by $p^\mu$, i.e. by numbers,
and in setting $p^2=m^2$, Eq.~(\ref{FNProj}) can be cast into the form
\beq
P^{{{15}\over 4}}   (\p)  &=& -{1\over 3 }\left[ 
- {1\over 2}S^{\mu\nu}S_{\mu\nu} +  {1\over m^2}
S_{\mu\nu}p^\mu S^{\rho\nu }p_\rho
+ {3\over 4}  \left( 1_4\otimes 1_4\right)  \right]\, ,\nonumber\\
P^{{{3}\over 4}}   (\p)  &=& +{1\over 3 }\left[ 
- {1\over 2}S^{\mu\nu}S_{\mu\nu}+{1\over m^2}
S_{\mu\nu}p^\mu S^{\rho\nu }p_\rho  +
{{15} \over 4}\left(1_4\otimes 1_4 \right) \right]\, ,\nonumber\\
\label{FNProj_const}
\eeq
Recall that the $\tau^+_1= {{15}\over 4}$ sector of $\Psi (\p )$ is a 
parity singlet.

\subsection{Propagators and Lagrangians for $\widetilde{W}^2$ invariant
subspaces}

In having favored the 16 dimensional vector column space $\Psi (\p ) $ 
over $\psi_\mu (\p )$, we gained the
advantage that the Dirac equation has been automatically accounted
for by means of the definition of the
Lorentz generators within $\Psi (\p ) $,
i.e. through the second term on the rhs  in Eq.~(\ref{psi_mu_gen}).
In so doing, one arrives at a wave equation for  
the $\tau^+_1 =15/4 $ sector of the four-vector-- spinor
that is quadratic in the momenta and reads 
\beq
-{1\over {3}}
{\Big[} - {1\over 2} S_{\mu\nu }S^{\mu\nu }
{{p^2}\over m^2}\, 
&+&S_{\mu\nu}p^\mu S^{\sigma\nu } {{ p_\sigma }\over m^2}\,  
+ {3\over 4} \left(1_4\otimes 1_4\right) {\Big]}
\Psi^{{{15}\over 4}} (\p ) =\Psi^{{{15}\over 4}} (\p )\, .\nonumber\\
\label{W2_Psimu_16}
\eeq
The propagator associated with  
Eq.~(\ref{W2_Psimu_16}) is now deduced as the following 
$16\times 16$  matrix:
\begin{eqnarray}
{\mathcal S}^{{{15}\over 4 }}(\p) &= &
 { {\widetilde{P}^{{ {15}\over 4}} (\p) }\over 
{p^2-m^2}}\, =  {{ 2S^{\mu\nu}S_{\mu\nu}- {4\over m^2}
S_{\mu\nu}p^\mu S^{\rho\nu }p_\rho
- 3\,\,  \left(1_4\otimes 1_4\right)  }\over 
{12 (p^2-m^2)} 
}\, .\nonumber\\ 
\label{FNProp}
\end{eqnarray} 
It directly verifies that Eq.~(\ref{W2_Psimu_16}) is obtained from the 
following Lagrangian
\begin{eqnarray}
{\mathcal L}^{{{15}\over 4}} &= & \bar{\Psi}^{{{15}\over 4}}(\p )\, 
 {\Big[} 
2m^2 S^{\mu\nu} S_{\mu\nu}  -
4 S_{\mu\nu} p^\mu S^{\rho\nu }p_\rho {\Big]}
 \Psi^{{{15}\over 4}} (\p )\nonumber\\
& -&  15  m^2 \, \left(1_4\otimes 1_4\right) 
\bar{\Psi}^{ { {15}\over 4} } (\p )\Psi^{{{15}\over 4}} (\p ) \, ,\nonumber\\
\bar\Psi (\p )\Psi (\p )&=&
{\Big[}A_\mu (\p ) g^{\mu\nu}\otimes 
\bar \psi (\p ){\Big]}\, 
{\Big[} \psi (\p )
\otimes A_\nu (\p ){\Big]}\, .  
\label{lagr}
\end{eqnarray}

\subsection{Energy-momentum dispersion relations
for  $\widetilde{W}^2$ invariant subspaces
in the presence of  electromagnetic fields }
 
In Ref.~[50] we studied the energy--momentum dispersion relation 
of Eq.~(\ref{W2_Psimu_16}) in the presence of
a simple magnetic field 
oriented along the $z$ axis, here denoted by $B_z$. We also took for
the sake of simplicity of the calculation the $z$ axis along $\p $. 
With the help of the symbolic code Mathematica we
then calculated  the appropriate determinant and, in nullifying it,
found the energy-momentum dispersion relation to be
\beq
E^2=(p_z -e  B_z )^2 + m^2\, ,
\label{gauge_disp}
\eeq
and therefore free from the Velo-Zwanziger problem 
of complex energy in the background of a magnetic field.
The associated  interacting propagators can now be obtained 
in the standard way by replacing $/\!\!\!p$ through
$/\!\!\!\pi :=(p^\mu-eA^\mu)\gamma_\mu $. 
In having done so, we have produced a 
{\it  pathology-free  propagating $\tau^+_1 ={{15}\over 4} $ sector \/}
in the presence of an electromagnetic field.
In this it was possible to avoid the 
Velo-Zwanziger problem.

The $\tau_1^+={{15}\over 4}$ sector corresponds to the ``has--been''
spin-${3\over 2}$ piece of $\psi_\mu (\0 )\simeq \Psi (\0 )$ at rest, 
which, after boosting,
ceases to be $\C^2$ eigenstate.
{} For a particle in the background of a magnetic field
we therefore obtained a  {\it pathology-free} covariant
description.

Admittedly, one has not produced pure-spin propagators.
Nonetheless, a formalism was created that at least 
allows for the covariant description of a propagating 
``has-been'' spin-${3\over 2}$ piece of the vector spinor. 
It may be useful  for the covariant description of
the $\Delta (1232)$ state, although the $\Delta (1232)$ resonance,
strictly speaking, does not belong to $\psi_\mu (\p ) $,
unless, some still undiscovered $\Delta $ state of spin-${1\over 2}^+$ 
happens to lie hidden in the vicinity of 1200 MeV. 

The trouble with the subsidiary conditions of the
Rarita-Schwinger framework has been that after gauging, 
Eq.~(\ref{projs1}) does no longer reduce to 
$\gamma^\mu \psi_\mu (\p )  =0$ for the $\tau_1^+={{15}\over 4}$
sector. In using the full covariant
projector, it was possible to circumvent the classical
Velo-Zwanziger problem in the covariant description of ``has--been''  
spin-${3\over 2}$ 
field.
As long as the tensors $\psi_{\mu_1...\mu_K} (\p )$ are totally 
symmetric,
the resolution of the problems related to any one of the indices
automatically applies to all of them.
Finally, that we worked in the sixteen dimensional vector column
space of $\Psi (\p ) $ rather than in $\psi_\mu (\p )$ did not 
restrict generality at all. Indeed, in order to construct vertices 
between $\Psi^{{15}\over 4} (\p  )$ and, say, the nucleon-pion system, 
it is sufficient to introduce the sixteen dimensional nucleon-pion
vector $\Psi^{\pi N} (\p ')=q^\pi _\mu\otimes \psi_N (\p^N )$ with 
$p'_\mu =q^\pi _\mu +p^N_\mu $, 
and $q^\pi _\mu$, and $p^N_\mu $ standing in turn for pion- and
nucleon momentum, and consider its overlap with 
$\Psi^{{15}\over 4} (\p )$.
In one of the possibilities, one finds
\beq
\bar \Psi^{{15}\over 4} (\p )\, 
\left( 1_4\otimes \gamma_\mu\gamma_5\right)\,
 \Psi^{\pi N} (\p ')
\eeq 
Vertices will be presented elsewhere.

\section{Perspectives} 
I reviewed the classification scheme for (non-strange) baryon spectra
in terms of $\left( {K\over 2},{K\over 2}\right)\otimes 
\Big[\left( {1\over 2},0\right)\oplus \left( 0,{1\over 2}\right) \Big]$
``barbed'' states (espinons),
 be they the non relativistic O(4) rotational and vibrational modes 
of a quark-diquark string, or, their covariant counterparts 
$\psi_{\mu_1...\mu_K}(\p )$ (up to form factors).

I briefly stressed on recent  QCD analyzes hinting on
a quark-diquark structure of the ground state baryons.
I argued that  a quark-(scalar diquark) string gives rise to natural
parity ``espinons'', while a quark-(pseudoscalar diquark) gives rise
to unnatural parity levels. The nucleon ground state and the
first ``espinon'' $\psi_\mu (\p ) $ were shown to belong to a 
quark-(scalar diquark) configuration,
while $\psi_{\mu_1...\mu_3}(\p )$, and 
$\psi_{\mu_1...\mu_5} (\p )$ belonged to
a quark- (pseudoscalar diquark) configuration.
Despite that the $\Delta (1232)$ ground state is well described by means of a
quark--(axial-vector) diquark string, in its excited states,
the angular momentum of the diquark  seems to change. 
I formulated chiral symmetry in baryon spectra in terms of 
$\psi_{\mu_1...\mu_K}(\p )$ and left the decision for
the TJNAF ``missing resonance'' program. 
Finding even one of the natural parities ``espinons'' 
$\psi_{\mu_1\mu_2}(\p )$, or,
$\psi_{\mu_1...\mu_4} (\p )$ would speak in favor
of chiral symmetry in the manifest Wigner-Weyl mode there.

As to the present status of affairs, I argued that one does not observe
genuine parity doublets, i. e. states having equal spins but
residing in {\it different\/} Fock spaces, i.e. spaces built on
top of vacua distinct through  opposite parities.
The occurrence of exactly $K$ states of equal spins and opposite
parities in same mass region is nothing but
an artifact of the parity degeneracy of the $K$ invariant subspaces
of the squared Pauli-Lubanski vector, $W^2$, in 
$\psi_{\mu_1...\mu_K} (\p )$,
a state residing as a whole, 
with all its ``has--been'' lower spin components in same Fock
space, i.e. built on top of same vacuum of either positive 
(scalar diquark), or, negative (pseudoscalar diquark) parities.
Within this context, I also clearly justified existence of 
parity singlet states as the $W^2$ sectors of the highest 
absolute value.

I reviewed  properties of the squared Pauli-Lubanski vector
$W^2$ in the direct products space  
$\left({1\over 2},{1\over 2}\right)
\otimes \Big[\left({1\over 2},0\right)\oplus \left(0,{1\over 2}\right)\Big]$,
which corresponds to the well known four-vector-spinor $\psi_\mu (\p )$.
The squared Pauli Lubanski vector was shown to
split $\psi_\mu (\p ) $ in two invariant subspaces
according to  $(W^2)^\delta_\eta  \,\psi^{\tau_l^\pm}_\delta (\p ) =
-\tau^\pm_l \,  m^2 \psi^{\tau_l^\pm}_\eta (\p )  $ with  $l=1,0$ and 
$\tau^+_1 ={{15}\over 4}$, $\tau^-_1=3/4$, and 
$\tau^\pm_0={{3}\over 4}$, respectively.
I  outlined construction of  covariant projectors and 
onto these subspaces, focused on the parity  non-degenerate
$W^2$ invariant subspace ( the has-been spin-${3\over 2}$ at rest) 
and showed that it is favorably described by a wave equation  that
does not suffer the Velo-Zwanziger problem of complex energies
in the presence of a magnetic field.
In this way a formalism was created for a covariant and pathology-free 
description of ``has--been'' higher-spin states at rest 
to the cost of giving up the demand on Lagrangians linear 
in the momenta.

\section*{Acknowledgments}
Dharam Vir Ahluwalia's invaluable drive
in delving jointly the {\it ab initio\/} constructs  
of the product representation spaces with all their 
mischievous traps is highly appreciated. 
Marcos Moshinsky and Yuri Smirnov brought their
unique expertise into uncovering the quark-diquark dynamics
behind the resonance clusters. 

Work supported by Consejo Nacional de Ciencia y
Tecnolog\'ia (CONACyT, Mexico) under grant number 32067-E.

\section{References}

\noindent
$^1$E.\ P.\ Wigner, Ann.\ of Math.\ {\bf 40}, 149 (1939).

\noindent
$^2$  C. Itzykson and J.-B. Zuber, {\it Quantum Field Theory\/}
(McGraw-Hill, 1988), 4th printing.

\noindent
$^3$ S.\ Weinberg, 
``Feynman Rules for Any Spin'', Phys.\ Rev.\ {\bf 133}, B1318-B1332 (1964);
S.\ Weinberg, {\em The Quantum
Theory of Fields, Vol. I and III\/} (Cambridge University Press,
Cambridge, 1995 and 2000).

\noindent
$^4$R.\ N.\ Mohapatra and  P.\ B.\ Pal,
{\em Massive Neutrinos in Physics and Astrophysics}
(World Scientific, Singapore, 1991).

\noindent
$^5$
G. H\"ohler, in {\em Pion-Nucleon Scattering\/} 
(Springer Publishers, Heidelberg, 1983), Landolt-B\"ornstein
Vol. I/9b2, Ed. H. Schopper.

\noindent
$^6$ M.\ Kirchbach,  
``On the parity degeneracy of baryons'',
Mod.\ Phys.\ Lett. {\bf A12}, 2373-2386
(1997);  
``Lorentz covariant spin-grouping of baryon resonances'',
Few Body Syst.\ Suppl.\ {\bf 11}, 47-52 (1999).

\noindent
$^7$Y.\ S.\ Kim and M.\  Noz,
{\it Theory and Application of the Poincar\'e Group\/} 
 (D.\ Reidel, Dordrecht, 1986).

\noindent
$^8$D.\ V.\ Ahluwalia and  M.\ Kirchbach,
``(1/2,1/2) representation space-an {\it ab inito\/} construct''
 Mod.\ Phys.\ Lett.\ {\bf A16}, 1377-1383  (2001).

\noindent
$^9$
M.\ Veltman, ``From weak interactions to gravitation'',
Int.\ J.\ Mod.\ Phys.\ {\bf A15}, 4557-4573 (2000).

\noindent
$^{10}$D.\ V.\ Ahluwalia, N.\ Dadhich, M.\ Kirchbach,
``On the spin of gravitational bosons'',
Int.\ J.\ Mod.\ Phys.\ {\bf D11}. 1621-1634 (2002).

\noindent
$^{11}$
M.\ Kirchbach and D.\ V.\ Ahluwalia,
``Spacetime structure of massive gravitino,''
Phys.\ Lett.\ {\bf B529}, 124-131 (2002).

\noindent
$^{12}$ M. Napsuciale, ``Principle of indistigushability and 
equations of motion for particles with spin'',
{\tt hep-ph/0212068}.

\noindent
$^{13}$ M.\ Kirchbach, M.\ Moshinsky, and Yu.\ F.\ Smirnov,
``Baryons in O(4) and vibron model'',
Phys.\ Rev.\ {\bf D64 }, 114005 (2001).

\noindent
$^{14}$ D. Singleton, 
``Electromagnetic answer to the nucleon spin puzzle'', 
Phys.\ Lett. {\bf B427}, 155-160 (1998).

\noindent
$^{15}$ D. Singleton {\it et al.}, contribution to this proceedings.

\noindent
$^{16}$ X. Ji, 
``Gauge invariant decomposition of nucleon spin and its spin-off'',
 Phys.\ Rev.\ Lett.\  {\bf 78}, 610-613 (1997); \\
X. Ji,
``How much of the nucleon spin is carried by glue?'', 
 Phys.\ Rev.\ Lett.\  {\bf 79}, 1255-1228 (1997). 

\noindent
$^{17}$ G.\ Savvidy,
``Gonihedric string equation.2'',
 Phys.\ Lett.\ {\bf B438}, 69-79 (1998).

\noindent
$^{18}$
W.\ Rarita and J.\ Schwinger, 
``On a theory of particles with half integral spin'',
Phys.\ Rev.\ {\bf 60}, 61 (1941).

\noindent
$^{19}$ G.\ Velo and D.\ Zwanziger, 
``Propagation and quantization of Rarita-Schwinger waves 
in an external electromagnetic potential'',
 Phys.\ Rev.\ {\bf 186}, 1337-1342 (1969). 

\noindent
$^{20}$ K.\ Johnson and E.\ C.\ G.\ Sudarshan,
``Inconsistency of local field theory of charged spin-3.2 particles'', 
 Ann.\ Phys.\ (N.\ Y.\ ){\bf 13}, 126-145 (1961)

\noindent
$^{21}$ V.\ Burkert, 
``Status of the N* program at Jefferson Lab'',
{\tt hep-ph/0210321}.

\noindent
$^{22}$
Particle Data Group, Eur.\ Phys.\ J.\,  {\bf C15}, 1 (2000).

\noindent
$^{23}$ M.\ Kirchbach,
``Degeneracy symmetry of baryon spectra'',
 Nucl.\ Phys.\ {\bf A689}, 157c-166c (2001).

\noindent
$^{24}$ R.\ Dashen,
``Chiral SU(3)XSU(3) as a symmetry of the strong interactions'',
 Phys. Rev. {\bf 183}, 1245-1260 (1969).

\noindent
$^{25}$F.\ D\"onau and H.\ Reinhardt,
{\it Spectroscopy of rotating bags\/}, 
preprint NBI-84-46 (1984) (unpublished).

\noindent
$^{26}$F.\ Iachello, `` Parity doubling in baryons and its
relevance to hadronic structure'', Phys.\ Rev.\ Lett.\ 
{\bf 62}, 2440-2443 (1989).

\noindent
$^{27}$ D. Robson, `` Comment on ``Parity doubling in baryons and its
relevance to hadronic structure'', 
Phys.\ Rev.\ Lett.\ {\bf 63}, 890-1891 (1989).

\noindent
$^{28}$ M.\ Kirchbach and  F.\ Beck 
{\it``Chiral Symmetry- the field  theoretical 
abstraction of the  enantiomorphism of matter''\/},
Preprint TH Darmstadt, IKDA 91/15\, .

\noindent
$^{29}$ M. Kirchbach, ``Elements of chiral symmetry'',
Czech.\ J.\ Phys.\ {\bf 43}, 319-340 (1993). 

\noindent
$^{30}$D.\ O.\ Riska, 
in  {\it Particle Physics. Proceedings 7th Jorge Andre  Swieca Summer School,
Campos do Jordao, Sao Paolo, Brazil, January 10-23, 1993\/},  
eds. O.\ J.\ P.\ Eboli and V.\ O.\ Rivelles
(World Scientific, Singapore, 1994). 

\noindent
$^{31}$L.\  Glozman and  D.\ O.\  Riska,
``The spectrum of the nucleons ans the strange hyperons and chiral dynamics'',
Phys.\ Rep.\ {\bf 268}, 263-303 (1996).

\noindent
$^{32}$ T.\ D.\ Cohen and L.\ Ya. Glozman, 
``Chiral multiplets versus parity doublets in highly excited baryons'',
Phys. Rev. {\bf D65}, 016006, 2002.

\noindent
$^{33}$  D.\ Jido, M.\ Oka, and A.\ Hosaka,
``Chiral symmetry of baryons'', Progr.\ Theor.\ Phys.\ 
{\bf 106}, 873-908 (2001).

\noindent
$^{34}$A.\ Hosaka, D.\ Jido, Y.\ Nemoto, and M.\ Oka,
``Chiral symmetry for baryons'',
             Nucl.\ Phys.\ {\bf A663}, 707-710 (2000).

\noindent
$^{35}$U.\ Lohring and B.\ Metsch, 
``The light baryon spectrum in a relativistic quark model with instanton 
induced quark forces: the strange baryon spectrum'' 
Eur.\ Phys.\ J.\ {\bf A10}, 447-486 (2001). 

\noindent
$^{36}$ R.\ Bijker, F.\ Iachello, and A.\ Leviatan,
``Electromagnetic form factors in a collective model of the nucleon'', 
                   Phys.\ Rev.\ {\bf C54}, 1935-1953 (1996).\\
                   R.\ Bijker, F.\ Iachello, and A.\ Leviatan,
``Algebraic models of hadron structure. 1.Nonstrange baryons','
                   Ann.\ of Phys.\ {\bf 236}, 69-116 (1994)
                
\noindent
$^{37}$ Proc. Int. Conf. {\it Diquarks 3}, Torino,
                    Oct. 28-30 (1996), eds. M.\ Anselmino and 
                    E.\ Predazzi, (World Scientific).

\noindent
$^{38}$M.\ Oettel, R.\ Alkofer, and  L.\ von Smekal 
`` Nucleon properties in the covariant quark diquark model,
Eur.\ Phys.\ J. {\bf A8}, 553-566 (2000). 

\noindent
$^{39}$ K.\ Kusaka, G.\ Piller, A.\ W.\ Thomas,  and
              A.\ G.\ Williams, 
`` Deep inelastic structure functions in a covariant spectator
model'', Phys.\ Rev.\  {\bf D55}, 5299-5308 (1997).

\noindent
$^{40}$F.\ Iachello, and R.\ D.\ Levin,
              {\it Algebraic Theory of Molecules\/}
               (Oxford Univ. Press, N.Y.) 1992.

\noindent
$^{41}$ J.\ P.\ Elliott and P.\ G.\ Dawber,
               {\it Symmetries in Physics\/} 
                 (The MacMillan Press Ltd, London, 1979).

\noindent
$^{42}$ M.\ Kirchbach, 
 ``Classifying reported and ``missing'' resonances according to their $P$ and
$C$ properties'', Int.\ J.\ Mod.\ Phys.\ {\bf A15}, 1435-1451 (2000).

\noindent
$^{43}$S.\ Capstick, T.\ S.\ H.\ Lee, W.\ Roberts, and A.\ Svar\'c,
``Evidence for the fourth P(11) resonance predicted by the
 constituent quark model'', 
Phys.\ Rev.\  {\bf C59}, 3002-3004 (1999).

\noindent
$^{44}$Yongseok Oh, A.\ I.\ Titov, and T.\ S.\ H.\ Lee,
``Higher and Missing Resonances in Omega Photoproduction'',               
{\tt nucl-th/0104046}

\noindent
$^{45}$ 
D. V. Ahluwalia, PhD Thesis, Texas A\&M University, 1991; 
{\it Dissertation Abstracts International B} {\bf 52}, 4792B (1992); 
D.~V. Ahluwalia and D.~J. Ernst, 
``(J,0) + (0,J) covariant spinors and causal 
propagators based on Weinberg formalism'',
Int.\ J.\ Mod.\ Phys.\ {\bf E2}, 397-422 
(1993).

\noindent
$^{46}$L.\ H.\ Ryder {\it Quantum field theory\/} 
(Cambridge Univ.\ Press, Cambridge,1987).

\noindent
$^{47}$M.\ Kirchbach, 
``Lorentz multiplet structure of baryon spectra and relativistic 
description'', Mod.\ Phys.\ Lett.\ {\bf A12}, 3177-3188 (1997). 

\noindent
$^{48}$E.\ Jenkins, 
``Chiral Lagrangians for baryons in the large 1/N(c) expansion'',
Phys.\ Rev.\ {\bf D53}, 2625-2644 (1996).

\noindent
$^{49}$ M.\ Kirchbach, D.\ V.\ Ahluwalia,
``Spacetime structure of massive gravitino'',
{\tt hep-ph/0210084  }

\noindent
$^{50}$ Gabriela Cabral, 
{\it Lagrangians for invariant subspaces of the squared 
Pauli-Lubanski vector, \/} thesis, FF-UAZ (in preparation).

\newpage
\vspace*{1truecm}
\begin{figure}[htb]
\vskip 5.0cm
\includegraphics{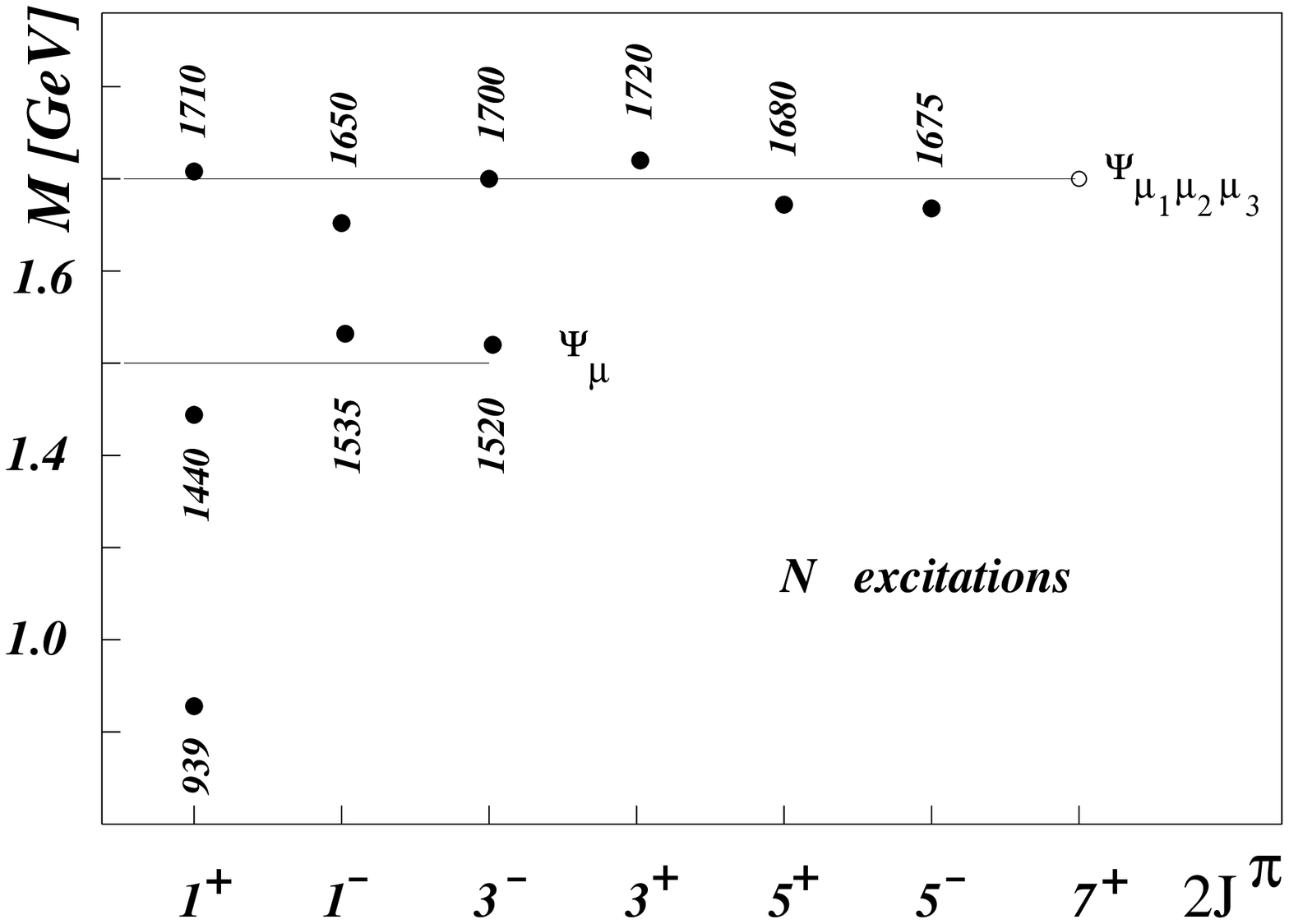}
\includegraphics{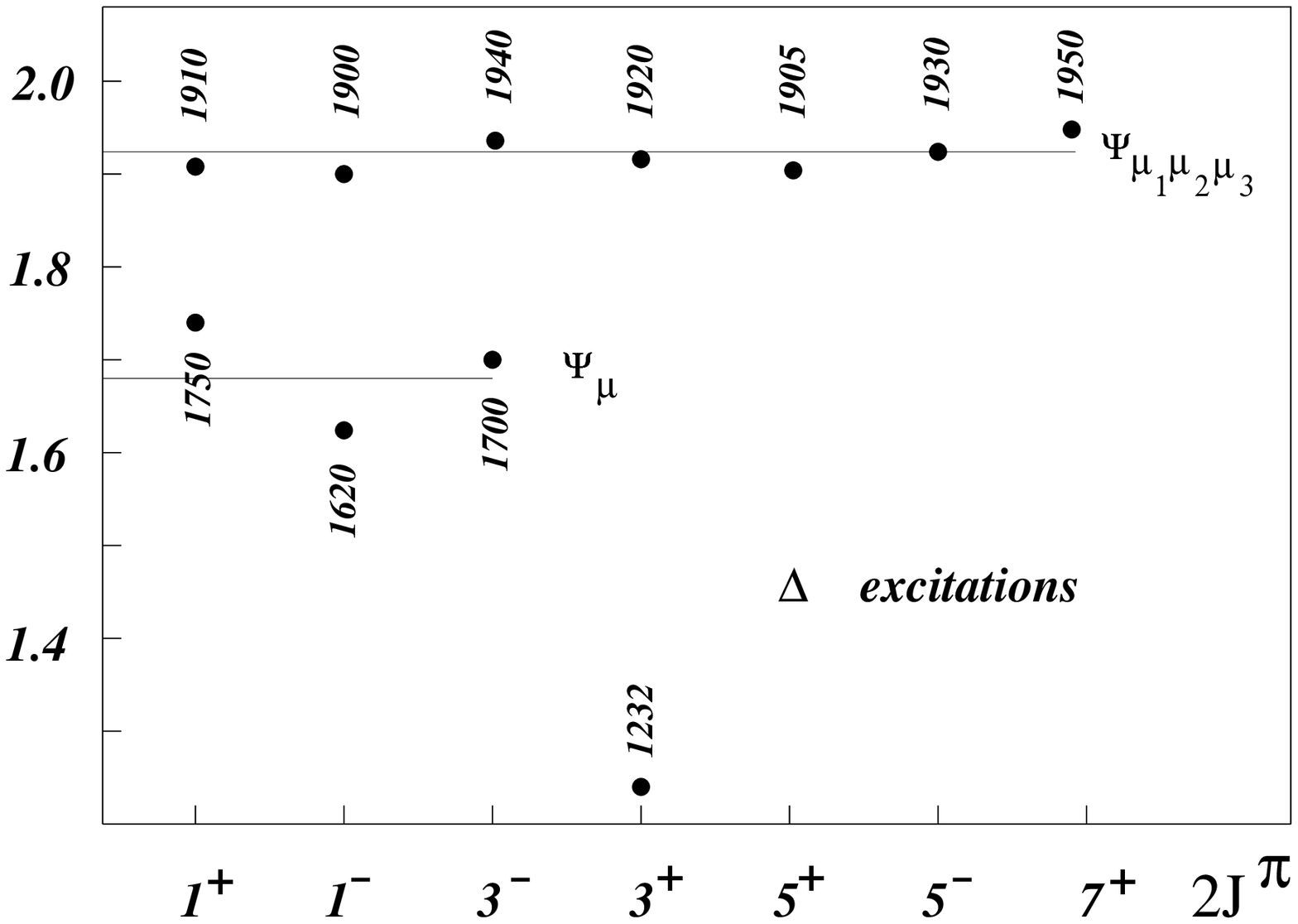}
\vspace{1.01cm}
{\small Figs.~1 and 2. 
Summary of the data on $N$ and $\Delta$ resonances.
The breaking of the mass degeneracy for each of the clusters
at about $5\%$ may in fact be an artifact of the data analysis,
as has been suggested by H\"ohler.$^5$
The filled circles represent known resonances, while 
the sole empty circle corresponds to a prediction. 
}
\end{figure}

\newpage

\begin{figure}[htbp]
\centerline{\psfig{figure=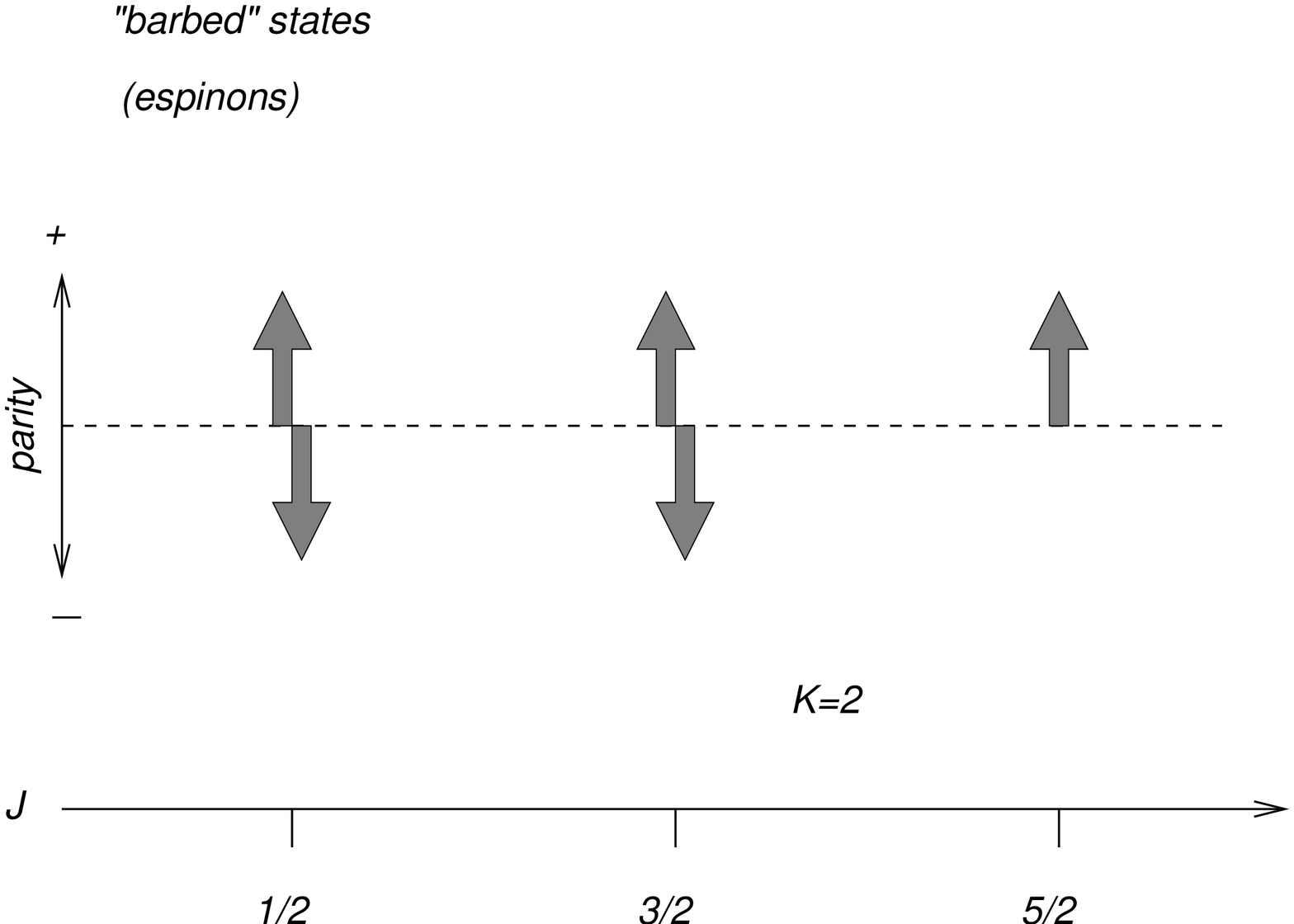,width=10cm}}
\vspace{0.1cm}
{\small Fig.~3
\hspace{0.2cm} 
$K$-excitation mode of a quark-diquark string: {\it barbed \/} states 
({\it espinons\/} ).
}
\label{fig:Esp_inon}
\end{figure}

\newpage
Table~ 1\\

\begin{table}[htbp]
\caption{
Predicted mass distribution of observed (obs), and
missing (miss) rovibron  clusters (in MeV) according to
Eq.~(34,35). The sign of $\eta $ in Eq.~(3) determines natural-
($\eta =+1$), or, unnatural ( $\eta =-1$) parity states.
{}All $\Delta $ excitations have been calculated
with  $f_2=40$ MeV rather than
with the nucleon value of $f_2=70$ MeV. 
The experimental mass averages of the resonances from a given
K--cluster have been labeled by ``exp''.
The nucleon and $\Delta $ ground state masses  
$M_N$ and $M_\Delta $ were taken to equal their experimental values. }
\vspace*{0.21truein}
\begin{tabular}{lccccccc}
\hline
~\\
K & sign $\eta $ & N$^{\mbox{obs} }$  & N$^{\mbox{exp}}$ &
 $ \Delta^{\mbox{obs}}$ & $\Delta^{\mbox{exp}}$ &  
   N$ ^{\mbox{miss}}$ & $\Delta^{\mbox{miss}}$  \\
\hline
~\\
0 & + &939 &939 & 1232& 1232 & &  \\
1&+  &1441 & 1498 & 1712 &1690 &         & \\
2&+  &     &      &      &     &  1612   & 1846 \\
3&-  &1764 &1689  & 1944 &1922 &         &     \\
4&+  &     &      &      &     &  1935   & 2048 \\
5&-  &2135 &2102  & 2165 &2276 &         &      \\
\hline
\end{tabular}
\end{table}

\newpage

Table~ 2\\

\begin{table}[htbp]
\caption{ Width of resonance clusters }
\begin{tabular}{lcc}
\hline
~\\
K&Resonance & width [in GeV]  \\
\hline
~\\
3 &$N\left( {1\over 2}^-;1650\right)$ & 0.15  \\
3 &$N\left( {1\over 2}^+;1710\right)$ & 0.10   \\
3 &$N\left( {3\over 2}^+;1720\right)$ & 0.15    \\
3 &$N\left( {3\over 2}^-;1700\right)$ & 0.15     \\
3 &$N\left( {5\over 2}^-;1675\right)$ & 0.15      \\
3 &$N\left( {5\over 2};^+1680\right)$ & 0.13       \\
~\\
\hline
5 &$N\left( {3\over 2}^+;1900\right)$ & 0.50\\
5 &$N\left( {5\over 2}^+;2000\right)$ &0.49\\
~\\\hline
\hline
\end{tabular}
\end{table}

\newpage

Table~ 3\\

\begin{table}[htbp]
\caption{ Helicity amplitudes of resonance clusters }
\begin{tabular}{lcccc}
\hline
~\\
K& parity of the spin-0 diquark  
&Resonance  &$A^p_{{1\over 2}}$ & $A_{{3\over 2}}^2$ 
[in $10^{-3}$GeV$^{-{1\over 2}}$]  \\
\hline
~\\
3 &- & $N\left( {1\over 2}^+;1710\right)$  &  9 $\pm$22 &  \\
3 &- & $N\left( {3\over 2}^+;1720\right)$  &  18$\pm$30 &-19$\pm$20   \\
3 &- & $N\left( {3\over 2}^-;1700\right)$  & -18$\pm$30 & -2$\pm$24    \\
3 &- & $N\left( {5\over 2}^-;1675\right)$  & 19 $\pm$8  & 15$\pm$9      \\
3 &- & $N\left( {5\over 2};^+1680\right)$  & -15$\pm$6  &133$\pm$12      \\
~\\
\hline
1 &+ &$N\left( {3\over 2}^-;1520\right) $ & -24$\pm$9 & 166$\pm$ 5\\
~\\\hline
\hline
\end{tabular}
\end{table}

\end{document}